\documentclass[preprint,12pt,authoryear]{elsart}  

\usepackage{natbib}

\usepackage{mathtools,mathrsfs,amsmath}
\usepackage{graphicx,color,caption}
\usepackage{subfigure}
\usepackage{bm}
\usepackage{amsfonts}
\usepackage{amssymb,amssymb,calligra,calrsfs}

\usepackage{etoolbox}
\usepackage{tikz}

\newrobustcmd*{\mysquare}[1]{\tikz{\filldraw[color=black,fill=#1] (0,0)
rectangle (0.2cm,0.2cm);}}

\newrobustcmd*{\mycircle}[1]{\tikz{\filldraw[color=black,fill=#1] (0,0) circle [radius=0.1cm];}}

\newrobustcmd*{\mytriangle}[1]{\tikz{\filldraw[color=black,fill=#1] (0,0) --
(0.2cm,0) -- (0.1cm,0.2cm) -- (0,0);}}

\newrobustcmd*{\myinvertedtriangle}[1]{\tikz{\filldraw[color=black,fill=#1] (0.1cm,0) --
(0,0.2cm) -- (0.2cm,0.2cm) -- (0.1cm,0);}}

\usepackage{lineno}

\newcommand{\norm}[1]{\vert{#1}\vert}		
\newcommand{\supofseq}[1]{\left\lVert\boldsymbol{#1}\right\rVert}	
\newcommand{\Tr}[1]{Tr\left({#1}\right)}	
\DeclarePairedDelimiter\ceil{\lceil}{\rceil}

\newtheorem{assumption}{Assumption}

\newtheorem{definition}{Definition}

\newtheorem{theorem}{Theorem}
\newtheorem{remark}{Remark}

\begin{document}
\nolinenumbers              
\begin{frontmatter}

\title{Adaptive polytopic estimation for nonlinear systems under bounded disturbances using moving horizon \thanksref{footnoteinfo}} 

\thanks[footnoteinfo]{The material in this paper was not presented at any conference.}

\author[sinc]{Nestor Deniz}\ead{ndeniz@sinc.unl.edu.ar},    
\author[sinc]{Marina Murillo}\ead{mmurillo@sinc.unl.edu.ar},
\author[sinc]{Guido Sanchez}\ead{gsanchez@sinc.unl.edu.ar},               
\author[sinc]{Leonardo Giovanini}\ead{lgiovanini@sinc.unl.edu.ar}  

\address[sinc]{Instituto de Investigacion en Senales, Sistemas e Inteligencia Computacional, sinc(\it{i}), UNL, CONICET, Ciudad Universitaria UNL, 4to piso FICH, (S3000) Santa Fe, Argentina
}  


\begin{abstract}                          
This paper introduces an adaptive polytopic estimator design for nonlinear systems under bounded disturbances combining moving horizon and dual estimation techniques. It extends the moving horizon estimation results for \emph{LTI} systems to polytopic \emph{LPV} systems. The design and necessary conditions to guarantee the robust stability and convergence to the true state and parameters for the case of bounded disturbances and convergence to the true system and state are given for the vanishing disturbances.
\end{abstract}

\end{frontmatter}

\section{Introduction}      \label{Introduction}
Accurate information of states, parameters and disturbances is essential for effective real-time operation of any system. Many of its relevant variables are often not measurable or too expensive to measure on line. A cost effective approach is to employ estimation techniques to obtain the required information from measurements of other variables and a mathematical model of the system. 
Linear and nonlinear estimation have been an active researcher field during the past several decades \citep{patwardhan2012nonlinear}. Linear estimation methods use a simpler representation of the system and can provide acceptable performance only around an operating point and the steady state operational conditions. However, as nonlinearities in the system dynamics become dominants, the performance of linear approaches deteriorates and the estimation algorithms will not necessarily converge to an accurate solutions. Although optimal state estimation solutions for linear systems exists, nonlinear estimation algorithms suffer from generating near-optimal solutions. Consequently, research of nonlinear estimation and filtering problems remains a challenging research area. 

Countless studies have been conducted in the literature to address and analyse nonlinear estimation problems. These methods can be broadly categorized into \citep{patwardhan2012nonlinear}: i) linearization methods \citep{kushner1977probability}; ii) approximation methods \citep{benevs1981exact}; iii) Bayesian recursive methods \citep{doucet2001introduction}; iv) moment methods \citep{crisan1998convergence}; and v) higher dimensional nonlinear filter methods \citep{arasaratnam2009cubature}.  However, there exist some approaches that approximate the nonlinear behaviour of systems with a linear
parameter--varying (LPV) models  \citep{shamma1991guaranteed,shamma1993gain}.

LPV systems are linear systems with matrices depending on time-varying parameters that can evolve over wide operating ranges  \citep{apkarian1995self}. These parameters, called \textit{scheduling variables}, depend on exogenous signals that can be measured. When the bounds of these signals are known, the LPV model can be reformulated into a convex linear combination of linear time-invariant \citep{leith2000survey}. If the scheduling variables are functions of endogenous signals such as states, inputs or outputs of the system instead of exogenous signals, \emph{LPV} system describes a large class of nonlinear systems \citep{toth2011state}. The most common technique to obtain an LPV system is the polytopic approach, where the system depends affinely on a time-varying parameter vector that evolves within a polytopic set. In practical situations they could be inaccessible by the fact that scheduling variables are functions of the system states \citep{theilliol2011design}.

In the polytopic LPV observer design, trust full knowledge of the scheduling variables is of paramount importance, because this information is needed to design the observer. Many researchers have proposed solutions to this problem in the polytopic framework. LPV observers with unmeasured scheduling parameters can be designed using proportional observer \citep{ichalal2016auxiliary}, proportional-integral observer \citep{aouaouda2013multi}, generalized dynamic observer \citep{gao2016new,osorio2016new} and adaptive observer \citep{bezzaoucha2013nonlinear,bezzaoucha2018new} framework, respectively.

The main contribution of this paper is the design and analysis of a robust estimator for nonlinear systems under bounded disturbances combining quasi-\emph{LPV} models and dual estimation using a receding horizon framework. The proposed algorithm simultaneously estimates the mixing parameters and the states using a dual estimation approach within a multiple iteration scheme that improve the performance of the estimation at each sample. The conditions to guarantee the robust stability and a convergence to the true system and states for the case of vanishing disturbances are derived. To achieve these results is crucial that the prior weighting in the cost function and the length of the estimation horizon are properly chosen. The assumption on the prior weighting can be verified a prior design.
The paper is organized as follows: Section 2 introduces the notation, definitions and properties that will be used through the paper. Section 3 presents the main results and shows its connections with previous results. The stability and convergence to the true state in the dual iteration is discussed in the initial part of section. Then, the robust regional stability and convergence to the true state and parameters of the  estimator are analysed. In section 4 two simple examples are discussed to illustrate the concepts and to show the difference with the estate of the art. Finally, Section 5 presents conclusions.

\section{Preliminaries and setup}

\subsection{Notation}
Let $\mathbb{Z}_{\left[a,b \right]}$ denotes the set of integers in the interval $\left[ a,b \right] \subseteq \mathbb{R} \textnormal{, and } \mathbb{Z}_{\geq a}$ denotes the set of integers greater or equal to $a$. Boldface symbols denote sequences of finite or infinite length, i.e., $\boldsymbol{w} \coloneqq \{ w_{k_1}, \ldots, w_{k_2}  \} \textnormal{ for some } k_1, k_2 \in \mathbb{Z}_{\geq 0} \textnormal{ and } k_1 < k_2$, respectively. We denote $x_{j\vert k}$ as the element of the sequence $\boldsymbol{x}$ given at time $k \in \mathbb{Z}_{\geq 0} \textnormal{ and } j \in \left[k_1, k_2 \right]$. 
By $\norm{x}$ we denote the Euclidean norm of a vector $x \in \mathbb{R}^n$. Let $\supofseq{x} \coloneqq \sup_{k\in \mathbb{Z}_{\geq 0}} \norm{x_k} $ denote the supreme norm of the sequence $\boldsymbol{x} \textnormal{ and } \supofseq{x}_{\left[a, b \right]} \coloneqq \sup_{k\in \mathbb{Z}_{\left[a, b \right]}} \norm{x_k} $ . 
A function $\gamma : \mathbb{R}_{\geq 0} \rightarrow \mathbb{R}_{\geq 0}$ is of class $\mathcal{K}$ if $\gamma$ is continuous, strictly increasing and $\gamma \left(0\right) = 0$ . If $\gamma$ is also unbounded, it is of class $\mathcal{K}_\infty$. A function $\zeta : \mathbb{R}_{\geq 0} \rightarrow \mathbb{R}_{\geq 0}$ is of class $\mathcal{L}$ if $\zeta \left( k \right)$ is non increasing and $\lim_{k \rightarrow \infty} \zeta\left(k \right) = 0$. A function $\beta : \mathbb{R}_{\geq 0}\times\mathbb{Z}_{\geq 0} \rightarrow \mathbb{R}_{\geq 0}$ is of class $\mathcal{KL}$ if $\beta \left(\cdot,k \right)$ is of class $\mathcal{K}$ for each fixed $k \in \mathbb{Z}_{\geq 0}$, and $\beta \left(r,\cdot \right)$ of class $\mathcal{L}$ for each fixed $r \in \mathbb{R}_{\geq 0}$. 

The following inequalities hold for all $\beta \in \mathcal{KL}, \; \gamma \in \mathcal{K} \textnormal{ and } a_j \in \mathbb R_{\geq 0} \textnormal{ with } j \in \mathbb{Z}_{\left[1,n \right]}$
\begin{equation}		\label{property_1}
  \begin{split}
    \gamma\left(\sum_{j=1}^{n}a_i \right) \leq \sum_{j=1}^{n} \gamma\left (n\, a_i\right),&\quad 
    \beta\left(\sum_{j=1}^{n}a_i, k \right) \leq \sum_{j=1}^{n}\beta\left(n\, 
    a_i, k\right).
  \end{split}
\end{equation}
The preceding inequalities hold since $\max\{a_j \}$ is included in the sequence $\{ a_1, a_2, \ldots, a_n \}$ and $\mathcal{K}$ functions are non-negative strictly increasing functions.

\textbf{Bounded sequences:} A sequence $\boldsymbol{w}$ is bounded if $\supofseq{w}$ is finite. The set of bounded sequences $\boldsymbol{w}$ is denoted as $\mathcal{W}\left(w_{\max}\right) \coloneqq \{w : \boldsymbol{w} \leq w_{\max} \}$ for some $w_{\max} \in \mathbb{R}_{\geq 0}$.

\textbf{Convergent sequences:}
A bounded infinite sequence $\boldsymbol{w}$ is convergent if $\norm{w_k} \rightarrow 0$ as $k \rightarrow \infty$. Let us denote the set of convergent sequences 
\begin{equation*}
 \mathcal{C}_w \coloneqq \{ \boldsymbol{w} \in \mathcal{W}\left(w_{\max}\right) 
    \vert \boldsymbol{w}\text{ is convergent} \}.   
\end{equation*}
Analogously, the sequence \textbf{$\boldsymbol{v}$} and $\mathcal{C}_v$ can be defined in similar way.
    

\subsection{Problem statement} 
Let us consider a nonlinear discrete-time system with the following behaviour
\begin{equation}		\label{eq_nonlinsys}
  \begin{split}
	x_{k+1} =&  f\left(x_k, w_k, \right) \qquad x_0 =  \mathtt{x}_0, \: \forall k \in \mathbb{Z}_{\geq 0},\\
	  y_{k} =& h\left(x_k\right) + v_k
  \end{split}
\end{equation}
where $x_{k} \in \mathcal{X} \subset \mathbb{R}^{n_x}$ is the system state, $w_{k} \in \mathcal{W} \subset \mathbb{R}^{n_x}$ is the additive process disturbance, $y_k \in \mathcal{Y} \subset \mathbb{R}^{n_p}$ is the system measurements and $v_{k} \in \mathcal{V} \subset \mathbb{R}^{n_p}$ is the measurement noise.
The sets $\mathcal{X}, \mathcal{W}, \mathcal{Y}$ and $\mathcal{V}$ are known compact and convex with the null vector $\textbf{0}$ in their interior. In the following we assume that  $f:\mathbb{R}^{n_x}\times\mathbb{R}^{n_m}\rightarrow \mathbb{R}^{n_x}$ is at least $C^1$ and locally Lipschitz on $x_k$ and  $h:\mathbb{R}^{n_x}\rightarrow \mathbb{R}^{n_p}$ is continuous.
Finally, the solution of system \eqref{eq_nonlinsys} at time $k$ is denoted by $x(k,x_0, \boldsymbol{w}, \boldsymbol{d})$, with initial condition $\mathtt{x}_0$ and disturbance sequence $\boldsymbol{w}$. Furthermore, the initial conditions $x_0$ and $\alpha_0$ are unknown, but priors knowledge $\bar{x}_0$ and $\bar{\alpha}_0$ are assumed to be available and their errors are assumed to be bounded, i.e., $\bar{x}_0 \in \mathcal{X}_0 \coloneqq \left\{ \bar{x}_0 : |x_0 - \bar{x}_0| \leq  e_{x\,max} \right\}$ such that $\mathcal{X}_0  \subseteq \mathcal{X}$ and $\bar{\alpha}_0 \in \mathcal{A}_0 \coloneqq \left\{ \bar{\alpha}_0 : |\alpha_0 - \bar{\alpha}_0| \leq  e_{\alpha\,max} \right\}$ such that $\mathcal{A}_0  \subseteq \mathcal{A}$, respectively.

The solution of the estimation problem aims to find at time $k$ an estimate $\hat{x}_{k|k}$ of the current state $x_k$ using a moving horizon estimator (\emph{MHE}). At each sampling time $k$ the only information available are the previous $N$ measurements $\boldsymbol{y} \coloneqq \left\{ y_{k-N} , \dotsc , y_{k} \right\}$ and a matrix $G(x_k,w_k) \in \Omega(\mathcal{A})$, where $\mathcal{A}$ denotes a polytopic set of matrices
such that
\begin{equation}
    A_k = \sum_{i=1}^{q} \alpha_{i,k} A_i, \quad \sum_{i=1}^{q} \alpha_{i,k} C_i
\end{equation}
with $\mathcal{A}$ the unit simplex
\begin{equation}
  \mathcal{A} \coloneqq \left\{ \displaystyle\sum_{i=1}^{q} \alpha_{i,k} = 1, \alpha_{i,k} \geq 0 \right\}
\end{equation}
Then, any property ensured for the uncertain \emph{LPV} model 
\begin{equation}
  \begin{array}{rl}
    x_{k+1} =& \displaystyle \sum_{i=1}^{q} \alpha_{i,k} A_i x_{k} + w_{k} + d_{k} , \\
    y_{k} =& \displaystyle \sum_{i=1}^{q} \alpha_{i,k} C_i x_{k} + v_{k},
  \end{array}
\end{equation}
holds true also for the nonlinear system \eqref{eq_nonlinsys} \citep{angelis2003system}. 
Therefore, in this work we propose a moving horizon estimation algorithm to simultaneously estimate the state of the system $\hat{x}_{k \vert k}$ and the mixing parameter of the  \emph{LPV} model $\hat{\alpha}_{k \vert k}$. The optimization problem to be solved at each sampling time is the following
\begin{equation}        \label{nonconvex problem}
\begin{array}{c}
  \underset{\boldsymbol{\hat{x}_{k-N|k}}, \boldsymbol{\hat{w}}, \boldsymbol{\hat{d}}, \boldsymbol{\hat{\alpha}}_k, \boldsymbol{\hat{w}_{\alpha}} }{\operatorname{min}} 
  \begin{array}{rl}
    \Psi_{x,\alpha} \coloneqq & \Gamma_{k-N} \left(\hat{x}_{k-N\vert k} \right)  + \sum\limits_{j=k-N}^{k} \ell \left(  \hat{w}_{j\vert k},\hat{v}_{j\vert k},\hat{w}_{{\alpha}_{j\vert k}},\hat{d}_{j\vert k}\right) \\
    & + \Lambda_k( \hat{\alpha}_{k-N\vert k})    
  \end{array}  \\
  \text{s.t.}\left\{
   \begin{array}{l}
    \begin{array}{rll}
     {\hat{\alpha}_{j+1\vert k}} =& {\hat{\alpha}_{j\vert k}} + {\hat{w}_{{\alpha}_{j\vert k}}} \hspace{4cm} j \in \mathbb{Z}_{\left[k-N, k-1\right]}, &  \\
     {\hat{x}_{j+1|k}}   =& \sum_{i=1}^{q} {\hat{\alpha}_{i,k\vert k}} A_i{\hat{x}_{j\vert k}} + {\hat{w}_{j\vert k}}+{\hat{d}_{j\vert k}} , \\
     y_{j} =& \sum_{i=1}^{q} {\hat{\alpha}_{i,k \vert k}} C_i {\hat{x}_{j\vert k}} + \hat{v}_{j\vert k} \hspace{2cm} j \in \mathbb{Z}_{\left[k-N, k\right]},  \\
     \sum_{i=1}^{q} {\hat{\alpha}_{i,k\vert k}} =& 1,\\
     {\hat{\alpha}_{i,k\vert k}} \geq& 0 \hspace{6cm} i \in \mathbb{Z}_{\left[1, q\right]},     \\
   \end{array}                        \\
  \hat{x}_{j\vert k} \in \mathcal{X}, \  \hat{w}_{j\vert k} \in \mathcal{W},\hat{w}_{\alpha} \in \mathcal{W}_{\alpha}, \  \hat{v}_{j\vert k} \in \mathcal{V}, \ \hat{d}_{j\vert k} \in \mathcal{D}.
  \end{array} \right.
\end{array}
\end{equation}
where $\hat{x}_{j|k}$ is the optimal estimated, $\hat{w}_{j|k}$ is the optimal process noise estimate and $\hat{\alpha}_{j|k}$ is the optimal mixing parameter and $\hat{w}_{\alpha_{j|k}}$ the noise associated to it at sample $k-j \quad j = 0, 1, \dotsc, N$ based on measurements $y_{k-j}$ available at time $k$. The process noise $\boldsymbol{\hat{w}} \coloneqq \left\{ \hat{w}_{k-N-1|k}, \dotsc, \hat{w}_{k-1|k} \right\}$, the mixing parameters ${\alpha} \coloneqq \left[ \alpha_{1,k|k}, \dotsc, \alpha_{q,k|k} \right]^T$, $\boldsymbol{\hat{w}_{\alpha}} \coloneqq \left\{ \hat{w}_{\alpha_{1,k|k}}, \dotsc, \hat{w}_{\alpha_{q,k|k}} \right\}$ and $\hat{x}_{k-N|k}$ are the optimization variables. The stage cost $\ell \left(\hat{w}_{j\vert k},\hat{v}_{j\vert k},\hat{w}_{{\alpha}_{j\vert k}},\hat{d}_{j\vert k}\right)$ penalizes the estimated process noise sequence $\boldsymbol{\hat{w}}_{j|k}$ and the estimation residuals $\boldsymbol{\hat{v}}_{j|k} = y_j - h\left(\hat{x}_{j|k}\right)$, while $\Gamma_{k-N} \left(\hat{x}_{k-N|k}\right)$ and $\Lambda\left(\hat{\alpha}_{k-N|k}\right)$ are the prior weights that penalizes the prior estimates $\hat{x}_{k-N|k}$ and $\hat{\alpha}_{k-N|k}$.

The robust stability of estimator \eqref{nonconvex problem} can be achieved by combining a suitable choice of the stage cost $\ell\left(  \hat{w}_{j\vert k}, \hat{v}_{j\vert k}, \hat{d}_{j\vert k} \right)$ and the time--varying prior weights
\begin{equation}
  \begin{array}{rl}
    \Gamma_{k-N \vert k}\left( \hat{x}_{k-N\vert k} \right) =& \lvert \, \hat{x}_{k-N\vert k} - \bar{x}_{k-N} \, \rvert_{P^{-1}_{x,k-N\vert k}}, \\
    \Lambda_{k-N \vert k} \left(\hat{w}_{\alpha,k-N\vert k} \right) =& \lvert \, \hat{\alpha}_{k-N\vert k} - \bar{\alpha}_{k-N} \, \rvert_{P^{-1}_{\alpha,k-N\vert k}}, 
  \end{array}
\end{equation}
whose parameters $\left( P^{-1}_{x,k-N\vert k},\bar{x}_{k-N}, P^{-1}_{\alpha,k-N\vert k},\bar{\alpha}_{k-N} \right)$ 
are recursively updated using the information available at time $k$. In this approach, the prior weight matrix $P_{*,k-N\vert k}$ are given by \citep{sanchez2017adaptive}
\begin{equation}  \label{eq:updatePk}
  \begin{split}
  \epsilon_{k-N} & = y_{k-N}-\hat{y}_{k-N \vert k}, 	\\
         N_{*,k} & = \left[ 1 + \hat{*}_{k-N\vert k-1}^T\, P_{*,k-N-1} \hat{*}_{*,k-N\vert k-1} \right]\frac{\sigma}
             {\norm{\epsilon_{k-N}}_2^2} \\
   \theta_{*,k} & = 1 - \frac{1}{N_{*,k}},\\
        W_{*,k} & = \left[ I - \frac{P_{*,k-N-1} \hat{*}_{k-
           N\vert k-1} \hat{*}_{k-N\vert k-1}^T}{1 + \hat{*}_{k-N\vert k-1}^T P_{*,k-N-1} \hat{*}_{k-N\vert k-1}}
           \right] P_{*,k-N-1}, \\
      P_{*,k-N} & = \left\{ \begin{array}{ccc}
           \frac{1}{\theta_{*,k}} W_{*,k} & & \text{if } \frac{1}{\theta_{*,k}}
           \Tr{W_{*,k}} \leq c, \\
           W_{*,k} & & \text{otherwise},
     \end{array} \right.
  \end{split}
\end{equation}
where $* \coloneqq \left[x,\alpha \right]$, $\sigma, \, \sigma_w, \, c, \, \lambda \in R_{>0}, \, c>\lambda, \, P_{0}=\lambda I_{n\times n}$ and $\sigma \gg \sigma_w$, where $\sigma_w$ denotes the process noise variance. The prior knowledges of the window $\bar{x}_{k-N}$ and $\bar{\alpha}_{k-N}$ are updated using a smoothed estimate
\begin{equation}  \label{eq:updateXA}
  \begin{array}{rl}
         \bar{x}_{k-N} =& \hat{x}_{k-N\vert k-1}, \\
    \bar{\alpha}_{k-N} =& \hat{\alpha}_{k-N\vert k-1}.
  \end{array}
\end{equation}

\subsection{Dual estimation formulation}

The joint estimator simultaneously estimates states and mixing parameters. For systems with many parameters, augmenting the state vector can cause a significant increase to the state dimension. This may be problematic as the dimension of the state vector grows, the errors accumulate and the convexity of the optimization problem is lost. To overcome this problem, a dual estimation setup is introduced: the estimation problem \eqref{nonconvex problem} solves separately the state estimation problem (assuming that mixing parameters $\alpha$ remains constant) and the model identification problem (assuming that estimated states $\hat{x}_k$ remains constant) at each sampling time. The problems to be solved iteratively are 
\begin{equation}        \label{mhe_x problem}
\begin{array}{c}
  \underset{\boldsymbol{\hat{x}_{k-N|k}}, \boldsymbol{\hat{w}_x} }{\operatorname{min}} \Psi_x \coloneqq \Gamma_{k-N} \left(\hat{x}_{k-N\vert k} \right) + \sum\limits_{j=k-N}^{k} \ell \left(  \hat{w}_{j\vert k}, \hat{v}_{j\vert k}\right) \\
  \text{s.t.} \left\{
  \begin{array}{l}
        \begin{array}{rll}
        {\hat{x}_{k-N\vert k}} 	=& \bar{x}_{k-N} + {\hat{w}_{k-N\vert k}} &   \\
        {\hat{x}_{j+1|k}} 	 	=& \sum_{i=1}^{q} \alpha_{i,k} A_i {\hat{x}_{j\vert k}} + {\hat{w}_{x \; j\vert k}} & \quad j \in \mathbb{Z}_{\left[k-N, k-1\right]},        \\
        y_{j} =& \sum_{i=1}^{q} \alpha_{i,k} C_i {\hat{x}_{j\vert k}} + \hat{v}_{j\vert k} & \quad j \in \mathbb{Z}_{\left[k-N, k\right]},  \\
        \end{array}                                                         \\
        \hat{x}_{j|k} \in \mathcal{X}, \  \hat{w}_{j|k} \in \mathcal{W}, \  \hat{v}_{j|k} \in \mathcal{V}.
  \end{array} \right.
\end{array}
\end{equation}
where the decision variables are $\hat{x}_{k-N\vert k}$ and $\hat{w}_{k-j\vert k} \quad j\coloneqq 1,2, \dots , N$  
and
\begin{equation}        \label{mhe_alpha problem}
\begin{array}{c}
  \underset{\boldsymbol{\alpha},\boldsymbol{\hat{w}_\alpha}}{\operatorname{min}} \Psi_{\alpha} \coloneqq \Lambda_{k-N} \left(\hat{\alpha}_{k-N\vert k} \right) + \sum\limits_{j=k-N}^{k} \ell \left( \hat{d}_{j\vert k}, \hat{v}_{j\vert k},\hat{w}_{{\alpha}_{j\vert k}}\right)                                   \\
  \text{s.t.} \left\{
  \begin{array}{l}
    \begin{array}{rll}
      {\alpha_{k-N\vert k}} =& \bar{\alpha}_{k-N} + {\hat{w}_{{\alpha}_{k-N\vert k}}} &  \\
      {\alpha_{j+1\vert k}} =& {\alpha_{j\vert k}} + {\hat{w}_{{\alpha}_{j\vert k}}}  & \; j \in \mathbb{Z}_{\left[k-N, k-1\right]} \\
      x_{j+1|k} =& \sum_{i=1}^{q} {\alpha_{i,j\vert k}} A^i x_{j\vert k} + 
      {\hat{d}_{j\vert k}}, 	 \\
      y_{j} =& \sum_{i=1}^{q} {\alpha_{i,j\vert k}} C^i x_{j\vert k} + \hat{v}_{j\vert k} &\; j \in \mathbb{Z}_{\left[k-N, k\right]},\\
      \sum_{i=1}^{q}{\alpha_{i,j\vert k}} =& 1    \\
      {\alpha_{i,j\vert k}} \geq& 0 & j \in \mathbb{Z}_{\left[1, q\right]}
    \end{array} \\
    \hat{d}_{j|k} \in \mathcal{D},\hat{w}_{\alpha} \in \mathcal{W}_{\alpha}, \  \hat{v}_{j|k} \in \mathcal{V}.
  \end{array} \right.
\end{array}
\end{equation}
where the decision variables are $\hat{w}_{\alpha,k-j\vert k}$ $\quad j\coloneqq 0,1, \dots , N$ and $\quad \hat{\alpha}_{k-N\vert k}$. Problems \eqref{mhe_x problem} and \eqref{mhe_alpha problem} are solved iteratively several times for the same sampling-time. The main novelty of the proposed algorithm is that an improvement in the state estimation and model identification can be guaranteed for a certain number of iterations when some assumptions are fulfilled. Moreover, the number of iterations can be computed offline.

The sequence $P_{k \vert k } \quad k \geq 0$ is positive definite, it is decreasing in norm and it is bounded. The proof of these properties follows similar steps as in \cite{sanchez2017adaptive}. 

\begin{assumption}  \label{prior weighting assumption}
The prior weighting $\Gamma_{k-N}\left(\hat{x}_{k-N\vert k}\right)$ is a continuous function $ \mathbb{R}^n \rightarrow \mathbb{R}$ lower bounded by $\underline{\gamma}_p\left(\cdot\right) \in \mathcal{K}_\infty{}$ and upper bounded by $\bar{\gamma}_p\left(\cdot\right) \in \mathcal{K}_{\infty}$ such that
\begin{equation}		\label{prior weighting bounds}
  \underline{\gamma}_p\left(\norm{\hat{x}_{k-N\vert k} - \bar{x}_{k-  N}} \right) \leq \Gamma_{k-N}\left(\norm{\hat{x}_{k-N\vert k} - \bar{x}_{k-  N}} \right) 
  \leq \bar{\gamma}_p \left(\norm{\hat{x}_{k-N\vert k} - \bar{x}_{k-N}} \right)
\end{equation}
for all $\hat{x} \in \mathcal{X}$ and
\begin{equation}        \label{eq:prior weighting equivalence}
  \underline{\gamma}_p\left(r \right) \geq \underline{c}_p \, r^{a},
  \quad \bar{\gamma}_p\left(r \right) \leq \bar{c}_p \, r^{a}
\end{equation}
where $0 \leq \underline{c}_p \leq \bar{c}_p$ and $a \in R_{\geq 1}$. Moreover, if the arrival cost is updated using equation \eqref{eq:updatePk}, the bounds $\underline{\gamma}_p$ and $\bar{\gamma}_p$ are bounded by
\begin{equation}
  \underline{\gamma}_p\left(r \right) \geq \norm{P^{-1}_0} \, r^{a},
  \quad \overline{\gamma}_p\left(r \right) \leq \norm{P^{-1}_{\infty}} \, r^{a}
\end{equation}
\end{assumption}

\begin{definition}          \label{ioss_definition}
The system \eqref{eq_nonlinsys} is \textit{incrementally input/output-to-state stable} if there exist some functions $\beta \in \mathcal{KL}$ and $\gamma_1, \gamma_2 \in \mathcal{K}$ such that for every two initial states $z_1$, $z_2 \in \mathbb{R}^n$, and any two disturbances sequences $\boldsymbol{w_1}, \boldsymbol{w_2}$ the following holds for all $k \in \mathbb{Z}_{\geq 0}$:
\begin{equation}        \label{i_IOSS_properterty}
  \begin{array}{rl}
    \norm{x\left( k,z_1,\boldsymbol{w_1} \right) - x\left( k,z_2,\boldsymbol{w_2} \right)} \leq& \max \left \{ \beta\left( \norm{z_1 - z_2},k \right),\gamma_1\left( \supofseq{w_1 - w_2}_{\left[0,k-1 \right]} \right),\right. \\
    &\left. \gamma_2\left( \supofseq{y_1 - y_2}_{\left[0,k-1 \right]} \right) \right \} \\
   \leq& \beta\left( \norm{z_1 - z_2},k \right) + \gamma_1\left( \supofseq{w_1 - w_2}_{\left[0,k-1 \right]} \right) + \\
   & \gamma_2\left( \supofseq{y_1 - y_2}_{\left[0,k-1 \right]} \right)
  \end{array}
\end{equation}
\end{definition}
for all $k$ \citep{sontag1997output}.
%
\begin{assumption}      \label{assumption beta function ineq}
The function $\beta(r,s) \in \mathcal{KL}$ and satisfies the following inequality:
\begin{equation}		\label{beta_function_ineq}
  \beta(r,s) \leq c_{\beta}r^p s^{-q}
\end{equation}
for some $c_{\beta} \in \mathbb{R}_{\geq 0}$, $p \in \mathbb{R}_{\geq 0}$ and $q \in \mathbb{R}_{\geq 0}$ and $q\geq p$.
\end{assumption}

\begin{assumption}		\label{stage cost assumption}
The stage cost $\ell\left(\cdot \right) : \mathbb{R}^n \times \mathbb{R}^m \rightarrow \mathbb{R}$ is a continuous function bounded by $\underline{\gamma}_w, \underline{\gamma}_v, \bar{\gamma}_w, \bar{\gamma}_v$ $\in \mathcal{K}_\infty{}$ such that the following inequalities are satisfied $\forall w \in \mathcal{W} \textnormal{ and } v \in \mathcal{V}$
  \begin{equation}		\label{stage cost inequalities}
    \underline{\gamma}_w\left(\hat{w} \right) + \underline{\gamma}_v\left(\hat{v} \right) \leq \ell\left(\hat{w}, \hat{v} \right) \leq \bar{\gamma}_w\left(\hat{w} \right) + \bar{\gamma}_v\left(\hat{v} \right)
  \end{equation}
\end{assumption}

\noindent Functions $\gamma_1$ and $\gamma_2$ from Definition \ref{ioss_definition} are related with the bounds of stage cost $\bar{\gamma}_w, \underline{\gamma}_w, \bar{\gamma}_v$  and $\underline{\gamma}_v$ through the following inequalities
\begin{equation}		\label{gamma1 and gamma2 inequalities}
  \gamma_1\left( 6\underline{\gamma}_w^{-1} \left( r \right)\right) \leq c_1
  r^{b_1}, \;
  \gamma_2\left( 6\underline{\gamma}_v^{-1} \left( r \right)\right) \leq c_2
  r^{b_2}
\end{equation}
for $c_{1}, c_{2}, b_{1},  b_{2} > 0.$

%

\begin{assumption}  \label{assum:equivalencia funcion gamma_alpha}
The prior weighting $\Lambda_{k-N}\left(\hat{\alpha}_{k-N\vert k}\right)$ is a continuous function $ \mathbb{R}^q \rightarrow \mathbb{R}$ lower bounded by $\underline{\gamma}_{\Lambda}\left(\cdot\right) \in \mathcal{K}_\infty{}$ and upper bounded by $\overline{\gamma}_{\Lambda}\left(\cdot\right) \in \mathcal{K}_{\infty}$ such that
\begin{equation}    \label{eq:equivalencia funcion gamma_alpha}
  \begin{array}{rcl}
    \underline{\gamma}_{\Lambda}\left(\norm{\hat{\alpha}_{k-N}-\bar{\alpha}_{k-N}}\right) &\leq \Lambda_{k-N}\left(\norm{\hat{\alpha}_{k-N}-\bar{\alpha}_{k-N}}\right) \leq& \overline{\gamma}_{\Lambda}\left(\norm{\hat{\alpha}_{k-N}-\bar{\alpha}_{k-N}}\right)
  \end{array}
\end{equation}
where
\begin{equation}
    \begin{array}{rcl}
     \underline{\gamma}_{\Lambda}\left(\norm{\alpha_{k-N}-\bar{\alpha}_{k-N}}\right)&\geq& \underline{c}_{\Lambda}\norm{\alpha_{k-N}-\bar{\alpha}_{k-N}}^a  \\
    \overline{\gamma}_{\Lambda}\left(\norm{\alpha_{k-N}-\bar{\alpha}_{k-N}}\right) &\leq&  \overline{c}_{\Lambda}\norm{\alpha_{k-N}-\bar{\alpha}_{k-N}}^a
    \end{array}
\end{equation}
for some $\underline{c}_{\Lambda}\in\mathbb{R}_{\geq0}$, $\overline{c}_{\Lambda}\in\mathbb{R}_{\geq0}$, $a\in\mathbb{R}_{>0}$, $\overline{c}_{\Lambda}>\underline{c}_{\Lambda}$.
\end{assumption}

In this work, we claim that the proposed estimator holds the property of being robust
asymptotic stable, which is defined as follows.
\begin{definition}
Consider the system described by Equation \eqref{mhe_alpha problem} subject to disturbances $\boldsymbol{w} \in \mathcal{W}\left(w_{\max} \right)$ and $\boldsymbol{v} \in \mathcal{V}\left(v_{\max} \right)$ for $w_{\max} \in \mathbb{R}_{\geq 0} $, $v_{max} \in \mathbb{R}_{\geq 0}$ with prior estimate $\bar{x}_0 \in \mathcal{X}\left(e_{\max} \right)$ for $e_{\max} \in \mathbb{R}_{\geq 0}$. The moving horizon state estimator given by Equation \eqref{mhe_alpha problem} is robustly
asymptotically stable (RAS) if there exists functions $\Phi \in \mathcal{KL} $ and $\pi_w$, $\pi_v \in \mathcal{K}$ such that for all $x_0 \in \mathcal{X}$, all $\bar{x}_0 \in \mathcal{X}_0\left(e_{\max} \right)$, the following is satisfied for all $k \in \mathbb{Z}_{\geq 0}$
\begin{equation}        \label{RGAS def property}
  \norm{x_k - \hat{x}_k} \leq \Phi\left(\norm{x_0 - \bar{x}_0},k \right) + \pi_w\left(\supofseq{w}_{[0,k-1]} \right) + \pi_v\left(\supofseq{v}_{[0,k-1]} \right)
\end{equation}
\end{definition}
We want to show that if system \eqref{mhe_x problem} is i-IOSS and Assumptions \eqref{prior weighting assumption}, \eqref{assumption beta function ineq} and \eqref{stage cost assumption}  are fulfilled, then the proposed estimator with adaptive arrival cost is RGAS. Furthermore, if the process disturbance and measurement noise sequences are convergent (i.e., $\boldsymbol{w}, \boldsymbol{v} \in \mathcal{C}$), the estimation converges to the true state.

\section{Theoretical properties}
Now we are ready to derive the main results: \textit{i}) the stability of the iterated dual estimation and \textit{ii} the robust
asymptotic stability (\emph{RAS}) of the proposed  estimator with an estimation horizon $\mathcal{N}$ for nonlinear detectable systems under bounded disturbances.

\subsection{Stability of the dual estimation iterations}
\label{sec:stability of the fixed point iteration}
As stated formerly, problems \eqref{mhe_x problem}--\eqref{mhe_alpha problem} are solved sequentially within a dual estimation iteration for each sampling time.
In the following, we will state the conditions required to achieve effectively a decreasing behaviour of the costs inside the dual estimation iteration. 
We have now all the necessary ingredients to enunciate the first theorem,
\begin{theorem}         \label{minimum value of l}
The sequences of costs $\{ \Psi_x^1,\Psi_x^2,\ldots,\Psi_x^l \}$ and $\{\Psi_{\alpha}^1,\Psi_{\alpha}^2, \ldots,\Psi_{\alpha}^l \}$ generated by the dual estimation iteration are decreasing if the number of iterations $l$ satisfies
\begin{equation}
  l \geq \displaystyle\log_2\left( \frac{\displaystyle\epsilon \Psi_x^1 - \Gamma_{k-N}\left( \hat{x}^l_{k-N\vert k} \right)}{\Gamma_{k-N}\left( \hat{x}^m_{k-N\vert k} \right)} +1 \right) +1,
\end{equation}
where $\epsilon \in \mathbb{R}_{\geq 0}$ and
\begin{equation*}
  \Gamma_{k-N}\left( \hat{x}^m_{k-N\vert k} \right)\coloneqq  \underset{i \in \mathbb{Z}_{\left[ 1,l \right]}}{\operatorname{min}} \left\{ \Gamma_{k-N}\left( \hat{x}^i_{k-N\vert k} \right) \right\}.   
\end{equation*}
\end{theorem}

\textbf{Proof.}
Let us consider the sequence of costs $\Psi_x^i(\cdot)$ and $\Psi_{\alpha}^i(\cdot)$ $\forall i \in \mathbb{Z}_{>1}$ generated in the \textit{i}--th iteration of the optimization problems \eqref{mhe_x problem} and \eqref{mhe_alpha problem}. Due to the optimality of the solutions the following inequalities are satisfied
\begin{equation}        \label{decreasing costs}
  \begin{array}{rl}
    \Psi_x^1\left( \cdot \right) &\geq \Psi_x^2\left( \cdot \right) \geq, \ldots, \geq \Psi_x^l\left( \cdot \right), \\ \Psi_{\alpha}^1\left( \cdot \right) &\geq \Psi_{\alpha}^2\left( \cdot \right) \geq, \ldots, \geq \Psi_{\alpha}^l\left( \cdot \right).
  \end{array}
\end{equation}
Since any iteration takes into account both $\Psi^i_x\left( \cdot \right)$ and $\Psi^i_{\alpha}\left( \cdot \right)$, and due to the sequences are non increasing, we only need to prove the decreasing behaviour of only one of these sequences, let's say $\Psi^i_x\left( \cdot \right)$.

Defining the normalized cost 
\begin{equation}        \label{gl}
  g(k,i) \coloneqq \frac{\Psi_x^i\left( \cdot \right)}{\Psi_x^1\left( \cdot \right)} \quad \forall k,i \geq 1,
\end{equation}
the necessary and sufficient conditions to guarantee its decrement along the dual iteration can be obtained using the Gronwall inequality (see \cite{ames1997inequalities, holte2009discrete}). It states that given any three non-negative sequences $y_n, f_n$ and $g_n$ that satisfy
\begin{equation}        \label{gronwall lemma condition} 
  y_n \leq f_n + \displaystyle \sum_{k=0}^{n}g_k \; y_k \qquad \forall n > 0,
\end{equation}
they also verify
\begin{equation}        \label{gronwall lemma inequality}
  y_n \leq f_n + \displaystyle \sum_{k=0}^{n-1}f_k g_k \displaystyle \prod_{j=k+1}^{n-1}\left( 1+g_j \right) \qquad \forall n > 0.
\end{equation}
Taking the sequences of costs $y_i = \Psi^i_x\left( \cdot \right)$, of arrival-costs $f_i = \Gamma_{k-N}\left( \cdot \right)$ and the normalized costs $g_i = g\left( k,i \right)$ for $n=l$
, which verify \eqref{gronwall lemma condition},
Gronwall inequality \eqref{gronwall lemma inequality} can be written as follows
\begin{equation} 
  \Psi_x^l \leq \Gamma_{k-N}\left(\hat{x}^l_{k-N\vert k} \right) + \displaystyle\sum_{i=1}^{l-1} \Gamma_{k-N}\left( \hat{x}_{k-N\vert k}^i \right) g(k,i) \prod_{j = i+1}^{l-1} \left( 1 + g(k,j) \right).
\end{equation}
Dividing by $\Psi_x^1$ we obtain
\begin{equation*}         \label{Gronwall ineq.}
   g\left( k,l \right) \leq \frac{\Gamma_{k-N}\left(\hat{x}^l_{k-N\vert k} \right) + \displaystyle\sum_{i=1}^{l-1} \Gamma_{k-N}\left( \hat{x}_{k-N\vert k}^i \right) g(k,i) \prod_{j = i+1}^{l-1} \left( 1 + g(k,j) \right)}{\displaystyle\Psi^1_x} = \epsilon,
\end{equation*}
which leads to 
\begin{equation}        \label{ineq intermedia 1}
  \displaystyle\sum_{i=1}^{l-1} \Gamma_{k-N}\left( \hat{x}_{k-N\vert k}^i \right) g(k,i) \prod_{j = i+1}^{l-1} \left( 1 + g(k,j) \right) = \epsilon \Psi_x^1 - \displaystyle\Gamma_{k-N}\left(\hat{x}^l_{k-N\vert k} \right).
\end{equation}
Defining
\begin{equation*}
  \Gamma_{k-N}\left( \hat{x}^m_{k-N\vert k} \right)\coloneqq  \underset{i \in \mathbb{Z}_{\left[ 1,l \right]}}{\operatorname{min}} \left\{ \Gamma_{k-N}\left( \hat{x}^i_{k-N\vert k} \right) \right\}   
\end{equation*}
and recalling that $g\left( k,i\right)$ is a non-increasing sequence, $g\left( k,l \right)\leq g\left( k,i \right) \; \forall l \geq i$, equation \eqref{ineq intermedia 1} can be rewritten as follows
\begin{equation}        \label{ineq intermedia 2}
  \Gamma_{k-N}\left( \hat{x}^m_{k-N\vert k} \right)g\left( k,l \right)\displaystyle\sum_{i=1}^{l-1}\prod_{j=i+1}^{l-1}\left( 1+g\left( k,l \right)  \right) < \epsilon \Psi_x^1 - \displaystyle\Gamma_{k-N}\left(\hat{x}^l_{k-N\vert k} \right).
\end{equation}
Since
\begin{equation}        \label{geometric serie}
  \displaystyle\sum_{i=1}^{l-1}\prod_{j=i+1}^{l-1}\left( 1+g\left( k,l \right)  \right) = \displaystyle \sum_{i=0}^{l-2}\left( 1+g\left( k,l \right) \right)^i = \frac{\displaystyle\left( 1+g\left( k,l \right) \right)^{l-1} -1 }{\displaystyle g\left( k,l \right)},
\end{equation}
inequality \eqref{ineq intermedia 2} can be rewritten as follows
\begin{equation}        \label{ineq intermedia 3}
  \begin{array}{rcl}
    \Gamma_{k-N}\left( \hat{x}^m_{k-N\vert k} \right)g\left( k,l \right)\left(\frac{\left( 1+g\left( k,l \right) \right)^{l-1} -1 }{ g\left( k,l \right)}\right) &\stackrel{\eqref{geometric serie}}{<}&  \epsilon \Psi_x^1 - \Gamma_{k-N}\left(\hat{x}^l_{k-N\vert k} \right) \\
  \end{array}
\end{equation}
and finally $g\left(k,l\right)$ is bounded by
\begin{equation}        \label{g(k,l)}
    g\left( k,l \right) < \left( \displaystyle\frac{\epsilon \Psi_x^1 - \Gamma_{k-N}\left(\hat{x}^l_{k-N\vert k} \right)}{\Gamma_{k-N}\left( \hat{x}^m_{k-N\vert k} \right)} + 1 \right)^{\displaystyle\frac{1}{l-1}}-1 < 1.
\end{equation}
Selecting a $l$ large enough, we can guarantee the decrement of sequence $g\left( k,i \right)$ and cost function $\Psi_x^i$ within the dual estimation iteration. Solving for inequality \eqref{g(k,l)}, an upper bound for the number of iterations is given by
\begin{equation}        \label{value of l}
     l \geq \ceil*{\log_2\left( \displaystyle\frac{\epsilon \Psi_x^1 - \Gamma_{k-N}\left(\hat{x}^l_{k-N\vert k} \right)}{\Gamma_{k-N}\left( \hat{x}^m_{k-N\vert k} \right)} + 1 \right)} + 1.
\end{equation}
A conservative estimate of $l$ can be computed taking into account the worst case scenario
\begin{equation}        \label{value of l conservative 1}
     l \geq \displaystyle \ceil*{\log_2\left( \mathcal{E} N\left( \bar{\gamma}_w\left( \supofseq{w} \right) + \bar{\gamma}_v\left( \supofseq{v} \right) \right) + 1 \right)} + 1,
\end{equation}
where $\mathcal{E} \coloneqq \epsilon/\Gamma_{k-N}\left( \hat{x}^m_{k-N\vert k} \right)$.

Inequalities \eqref{value of l} and \eqref{value of l conservative 1} allow to compute the required value of $l$ to guarantee the costs decreasing within the dual estimation iteration. \qed

\begin{remark}
Note that for the noiseless case, only one iteration is needed after the transient due to the uncertainty in the initial condition has vanished.
\end{remark}

\subsection{Robust stability}

In the previous subsection it was shown that the sequence of cost decreases within the dual estimation iteration. At each sampling time, the model used by the estimator is replaced with the newly available until satisfied the stopping criteria. In the following paragraphs we will prove robust stability for the estimator under bounded disturbances and model uncertainty assuming that the system is \textit{i}-IOSS. Moreover, if the length $N$ of the horizon of the estimator is larger than a certain value $\mathcal{N}$ that can be computed offline, the number of iterations $l$ is chosen according to equations \eqref{value of l} and \eqref{value of l conservative 1}, the effects of uncertainty in the initial condition vanish, as well as the disturbances due to model uncertainty. Besides, in the absence of process and measurement noises, states and model converges to the true ones.

\begin{theorem}		\label{Theorem_1}
Consider an i-IOSS system \eqref{eq_nonlinsys} with disturbances $\boldsymbol{w} \in \mathcal{W}\left(w_{\max} \right)$, $\boldsymbol{v} \in \mathcal{V}\left(v_{\max} \right)$.  Assume that the arrival cost weight matrix of the MHE problem $\Gamma_{k-N}$ is updated using the adaptive algorithm \eqref{eq:updatePk}. Moreover, Assumptions \ref{prior weighting assumption}, \ref{assumption beta function ineq} and \ref{stage cost assumption} are fulfilled and initial condition $x_0$ and $\alpha_0$ are unknown, but prior estimates $\bar{x}_0 \in \mathcal{X}_0$ and $\bar{\alpha}_0 \in \mathcal{A}_0$ are available. Then, the MHE estimator resulting from  problems \eqref{mhe_x problem}-- \eqref{mhe_alpha problem} is $RAS$. 
\end{theorem}

\textbf{Proof.}
In order to proof stability for the estimator, we start comparing the costs of the first iteration
and the resulting estimated state $\hat{x}_{k-N|k}$ for sampling time $k$
\begin{equation*}
  \begin{array}{rl}
    \Psi\left( \hat{x}_{k-N\vert k}, \hat{\alpha}, \left\{ \hat{\boldsymbol{w}}_{j\vert k} \right\}, \left\{ \hat{\boldsymbol{d}}_{j\vert k} \right\} \right) 
    &= g\left(k,l\right)\Psi^1\left( \hat{x}_{k-N\vert k}^1, \hat{\alpha}^1, \left\{ \hat{\boldsymbol{w}}_{j\vert k}^1 \right\}, \left\{ \hat{\boldsymbol{d}}_{j\vert k}^1 \right\} \right) \\
    &\leq g\left(k,l\right)\Psi\left( x_{k-N}, \alpha, \left\{\boldsymbol{w}\right\}, \left\{\boldsymbol{0}\right\} \right)
  \end{array}
\end{equation*}
Note that if $\hat{\alpha}=\alpha$, the estimated model match with the system, therefore there is no model uncertainty, i.e. $\left\{\boldsymbol{d}\right\} = \left\{\boldsymbol{0}\right\}$. Replacing the sequence of estimated process noises $\left\{\hat{\boldsymbol{w}}\right\}$ by the true sequence $\left\{\boldsymbol{w}\right\}$, the only feasible solution is the true sequence of states $\left\{\boldsymbol{x}\right\}$, and due optimality the inequality is verified. 
By mean of Assumptions \ref{prior weighting assumption} - \ref{assum:equivalencia funcion gamma_alpha}, the cost $\Psi\left( \cdot \right)$ is bounded by
\begin{equation}            \label{eq:detectability}
  \begin{array}{rl}
    \underline{\gamma}_p\left(\norm{\hat{x}_{k-N\vert k}-\bar{x}_{k-N}}\right) + \underline{\gamma}_{\Lambda}\left(\norm{\hat{\alpha}_{k-N\vert k}-\bar{\alpha}_{k-N}}\right) + 
    N\underline{\gamma}_w\left(  \norm{\hat{w}_{j\vert k}}\right) + \quad\\ 
    \underline{\gamma}_{\alpha}\left(\norm{\hat{w}_{\alpha, j\vert k}}\right)+ 
    N\underline{\gamma}_v\left(\norm{\hat{v}_{j\vert k}}\right) + N\underline{\gamma}_d\left(\norm{\hat{d}_{j\vert k}}\right) \leq& \\
    \left(\overline{\gamma}_p\left(\norm{x_{k-N}-\bar{x}_{k-N}}\right) + \overline{\gamma}_{\Lambda}\left(\norm{\alpha_{k-N}-\bar{\alpha}_{k-N}}\right) + N\overline{\gamma}_w\left(  \|\boldsymbol{w}\|_{\left[k-N,k\right]}\right) +\right. \quad\\
    \left.N\overline{\gamma}_{\alpha}\left(  \|\boldsymbol{w}_{\alpha}\|_{\left[k-N,k\right]}\right) +
    N\overline{\gamma}_v\left(\|\boldsymbol{v}\|_{\left[k-N,k\right]}\right)\right)g\left(k,l\right) \leq& \\
    \overline{c}_p\norm{x_{k-N}-\bar{x}_{k-N}}^a g\left(k,l\right)+\overline{c}_{\Lambda}\norm{\alpha_{k-N}-\bar{\alpha}_{k-N}}g\left(k,l\right)+N\left(\overline{\gamma}_w\left(\|\boldsymbol{w}\|\right) +\right. \;\\
    \left.
    \overline{\gamma}_v\left(\|\boldsymbol{v}\|\right)+\overline{\gamma}_{\alpha}\left(\|\boldsymbol{w_{\alpha}}\|\right)
    \right)g\left(k,l\right).\;
  \end{array}
\end{equation}
Solving for $\norm{\hat{x}_{k-N\vert k}-\bar{x}_{k-N}}$ and using relations \eqref{property_1}, we can write
\begin{equation}        \label{eq:bound of arrival cost 1-iteration}
    \begin{array}{rl}
         \norm{\hat{x}_{k-N\vert k}-\bar{x}_{k-N}} \leq& \underline{\gamma}_p^{-1}\left(\overline{c}_p\norm{x_{k-N}-\bar{x}_{k-N}}^a g\left(k,l\right)\right) + \\ 
         & \underline{\gamma}_p^{-1}\left(5\overline{c}_{\Lambda}\norm{\alpha_{k-N}-\bar{\alpha}_{k-N}}g\left(k,l\right)\right) + \\
         & \underline{\gamma}_p^{-1}\left(5N\overline{\gamma}_{w}\left(\|\boldsymbol{w}\|_{\left[k-N,k\right]}\right)g\left(k,l\right)\right) + \\
         & \underline{\gamma}_p^{-1}\left(5N\overline{\gamma}_{\alpha}\left(\|\boldsymbol{w_{\alpha}}\|_{\left[k-N,k\right]}\right)g\left(k,l\right)\right) +\\
         & \underline{\gamma}_p^{-1}\left(5N\overline{\gamma}_{v}\left(\|\boldsymbol{v}\|_{\left[k-N,k\right]}\right)g\left(k,l\right)\right).
    \end{array}
\end{equation}
Using again Assumptions \ref{prior weighting assumption} - \ref{assum:equivalencia funcion gamma_alpha}, one can write
\begin{equation}        \label{eq:bound of arrival cost 1-iteration with upper bound of functions}
    \begin{array}{rl}
         \norm{\hat{x}_{k-N\vert k}-\bar{x}_{k-N}} \leq& \displaystyle \left(\frac{5\overline{c}_p\norm{x_{k-N}-\bar{x}_{k-N}}^a g\left(k,l\right)}{\underline{}{c}_p}\right)^{1/a} + \\ 
         & \displaystyle \left(\frac{5\overline{c}_{\Lambda}\norm{\alpha_{k-N}-\bar{\alpha}_{k-N}}^a g\left(k,l\right)}{\underline{c}_p}\right)^{1/a} + \\
         & \displaystyle \left(\frac{5N\overline{\gamma}_{w}\left(\|\boldsymbol{w}\|_{\left[k-N,k\right]}\right)g\left(k,l\right)}{\underline{c}_p}\right)^{1/a} + \\
         & \displaystyle \left(\frac{5N\overline{\gamma}_{\alpha}\left(\|\boldsymbol{w_{\alpha}}\|_{\left[k-N,k\right]}\right)g\left(k,l\right)}{\underline{c}_p}\right)^{1/a} + \\
         & \displaystyle \left(\frac{5N\overline{\gamma}_{v}\left(\|\boldsymbol{v}\|_{\left[k-N,k\right]}\right)g\left(k,l\right)}{\underline{c}_p}\right)^{1/a}
    \end{array}
\end{equation}

From now on, we will drop the superindex $l$. 
Using Definition 1, the estimation error at time $k$, given the error at initial conditions $(k=0)$, is bounded by
\begin{equation*}
  \norm{x_k - \hat{x}_{k\vert k}} 
%
  \leq \beta\left( \norm{x_0 - \hat{x}_{0\vert k}},k \right)+\gamma_1\left( \supofseq{w - \hat{w}}_{\left[0,k-1 \right]} \right) + 
  \gamma_2\left(\supofseq{v - \hat{v}}_{\left[0,k-1 \right]} \right),    
\end{equation*}
and, assuming that $k=N$, we have
\begin{equation}        \label{eq:estimation error l-th it.}
  \begin{array}{rl}
  \norm{x_k - \hat{x}_{k\vert k}} \leq&  \beta\left( \norm{x_{k-N} - \hat{x}_{k-N\vert k}},N \right)+\gamma_1\left( \supofseq{w - \hat{w}}_{\left[k-N,k-1 \right]} \right) + \\
  & \gamma_2\left( \supofseq{v - \hat{v}}_{\left[k-N,k-1 \right]} \right).
  \end{array}
\end{equation}
To found a bound for the estimation error we need to bounds for the terms of the right hand of \eqref{eq:estimation error l-th it.}. Let us start with the first term using inequalities \eqref{property_1} such that the effect of estimation error at the beginning of the estimation window is bounded by
\begin{equation}    \label{eq:partial bound of beta function}
    \begin{array}{rl}
         \beta\left( \norm{x_{k-N} - \hat{x}_{k-N\vert k}},N \right) 
        &= \beta\left(\norm{x_{k-N}-\bar{x}_{k-N}+\bar{x}_{k-N} - \hat{x}_{k-N\vert k}},N\right)\\
        &\leq \beta\left(\norm{x_{k-N}-\bar{x}_{k-N}}+\norm{\bar{x}_{k-N} - \hat{x}_{k-N\vert k}},N\right)\\
        &\leq \beta\left(\norm{x_{k-N}-\bar{x}_{k-N}}+\norm{\hat{x}_{k-N\vert k}-\bar{x}_{k-N}},N\right) \\
        & \leq \beta\left(2\norm{x_{k-N}-\bar{x}_{k-N}},N\right)+\beta\left(2\norm{\hat{x}_{k-N\vert k}-\bar{x}_{k-N}},N\right). 
    \end{array}
\end{equation}
Now, the first term of \eqref{eq:partial bound of beta function} can be rewritten using Assumption \ref{assumption beta function ineq}, and the second term with the use of \eqref{eq:bound of arrival cost 1-iteration with upper bound of functions}
\begin{equation}
  \begin{array}{rl}
    \beta\left(\norm{x_{k-N}-\hat{x}_{k-N\vert k}},N\right) 
    \leq & \displaystyle \frac{c_{\beta}2^p \norm{x_{k-N}-\bar{x}_{k-N}}^p}{N^q} + \\
    & \beta\left(\displaystyle \frac{2\,5^{1/a}\overline{c}_p^{1/a}\norm{x_{k-N}-\bar{x}_{k-N}}g\left(k,l\right)^{1/a}}{\underline{c}_p^{1/a}} \right.+\\
    & \left.\displaystyle \frac{2\,5^{1/a}\overline{c}_{\Lambda}^{1/a}\norm{\alpha_{k-N}-\bar{\alpha}_{k-N}}g\left(k,l\right)^{1/a}}{\underline{c}_p^{1/a}} \right.+\\
    & \left.\displaystyle \frac{2\,5^{1/a}N^{1/a}\overline{\gamma}_w\left(\|\boldsymbol{w}\|_{\left[k-N,k-1\right]}\right)^{1/a}g\left(k,l\right)^{1/a}}{\underline{c}_p^{1/a}} \right.+\\
    & \left.\displaystyle \frac{2\,5^{1/a}N^{1/a}\overline{\gamma}_v\left(\|\boldsymbol{v}\|_{\left[k-N,k-1\right]}\right)^{1/a}g\left(k,l\right)^{1/a}}{\underline{c}_p^{1/a}} \right.+\\
    & \left.\displaystyle \frac{2\,5^{1/a}N^{1/a}\overline{\gamma}_{\alpha}\left(\|\boldsymbol{w_{\alpha}}\|_{\left[k-N,k-1\right]}\right)^{1/a}g\left(k,l\right)^{1/a}}{\underline{c}_p^{1/a}}, N \right)\\
  \end{array}
\end{equation}
Using inequalities \eqref{property_1}
\begin{equation}
    \begin{array}{rl}
    \beta\left(\norm{x_{k-N}-\hat{x}_{k-N\vert k}},N\right)
         \leq & \displaystyle \frac{c_{\beta}2^p \norm{x_{k-N}-\bar{x}_{k-N}}^p}{N^q} + \\
    & \beta\left(\displaystyle \frac{10\,5^{1/a}\overline{c}_p^{1/a}\norm{x_{k-N}-\bar{x}_{k-N}}g\left(k,l\right)^{1/a}}{\underline{c}_p^{1/a}}, N\right)+\\
    & \beta\left(\displaystyle \frac{10\,5^{1/a}\overline{c}_{\Lambda}^{1/a}\norm{\alpha_{k-N}-\bar{\alpha}_{k-N}}g\left(k,l\right)^{1/a}}{\underline{c}_p^{1/a}}, N\right)+\\
    & \beta\left(\displaystyle \frac{10\,5^{1/a}N^{1/a}\overline{\gamma}_w\left(\|\boldsymbol{w}\|_{\left[k-N,k-1\right]}\right)^{1/a}g\left(k,l\right)^{1/a}}{\underline{c}_p^{1/a}}, N\right)+\\
    & \beta\left(\displaystyle \frac{10\,5^{1/a}N^{1/a}\overline{\gamma}_v\left(\|\boldsymbol{v}\|_{\left[k-N,k-1\right]}\right)^{1/a}g\left(k,l\right)^{1/a}}{\underline{c}_p^{1/a}}, N\right)+\\
    & \beta\left(\displaystyle \frac{10\,5^{1/a}N^{1/a}\overline{\gamma}_{\alpha}\left(\|\boldsymbol{w_{\alpha}}\|_{\left[k-N,k-1\right]}\right)^{1/a}g\left(k,l\right)^{1/a}}{\underline{c}_p^{1/a}}, N\right)+\\
    \end{array}
\end{equation}
Now, by mean of Assumption \ref{assumption beta function ineq}
\begin{equation}
    \begin{array}{rl}
    \beta\left(\norm{x_{k-N}-\hat{x}_{k-N\vert k}},N\right)
        \leq & \displaystyle \frac{c_{\beta}2^p \norm{x_{k-N}-\bar{x}_{k-N}}^p}{N^q} + \\
        & \displaystyle \frac{c_{\beta}10^p\,5^{p/a}\overline{c}_p^{p/a}\norm{x_{k-N}-\bar{x}_{k-N}}^p g\left(k,l\right)^{p/a}}{N^q\,\underline{c}_p^{p/a}}  +\\
        & \displaystyle \frac{c_{\beta}10^p\,5^{p/a}\overline{c}_{\Lambda}^{p/a}\norm{\alpha_{k-N}-\bar{\alpha}_{k-N}}^p g\left(k,l\right)^{p/a}}{N^q\,\underline{c}_p^{p/a}} +\\
        & \displaystyle \frac{c_{\beta}10^p\,5^{p/a}N^{1/a}\overline{\gamma}_w\left(\|\boldsymbol{w}\|_{\left[k-N,k-1\right]}\right)^{1/a}g\left(k,l\right)^{1/a}}{N^q\,\underline{c}_p^{p/a}} +\\
        & \displaystyle \frac{c_{\beta}10^p\,5^{p/a}N^{p/a}\overline{\gamma}_v\left(\|\boldsymbol{v}\|_{\left[k-N,k-1\right]}\right)^{p/a}g\left(k,l\right)^{p/a}}{N^q\,\underline{c}_p^{p/a}} +\\
        & \displaystyle \frac{c_{\beta}10^p\,5^{p/a}N^{p/a}\overline{\gamma}_{\alpha}\left(\|\boldsymbol{w_{\alpha}}\|_{\left[k-N,k-1\right]}\right)^{p/a}g\left(k,l\right)^{p/a}}{N^q\,\underline{c}_p^{p/a}}
    \end{array}
\end{equation}
Rearranging terms
\begin{equation}
\label{eq:bound of first term of i-IOSS}
    \begin{array}{rl}
    \beta\left(\norm{x_{k-N}-\hat{x}_{k-N\vert k}},N\right)
        \leq & \displaystyle \frac{\norm{x_{k-N}-\bar{x}_{k-N}}^p}{N^q}\left(
        \displaystyle c_{\beta}2^p + \frac{c_{\beta}10^p\,5^{p/a}\overline{c}_p^{p/a}g\left(k,l\right)^{p/a}}{\underline{c}_p^{p/a}}\right)  +\\
        & \displaystyle \frac{c_{\beta}10^p\,5^{p/a}\overline{c}_{\Lambda}^{p/a}\norm{\alpha_{k-N}-\bar{\alpha}_{k-N}}^p g\left(k,l\right)^{p/a}}{N^q\,\underline{c}_p^{p/a}} +\\
        & \displaystyle \frac{c_{\beta}10^p\,5^{p/a}N^{1/a}\overline{\gamma}_w\left(\|\boldsymbol{w}\|_{\left[k-N,k-1\right]}\right)^{1/a}g\left(k,l\right)^{1/a}}{N^q\,\underline{c}_p^{p/a}} +\\
        & \displaystyle \frac{c_{\beta}10^p\,5^{p/a}N^{p/a}\overline{\gamma}_v\left(\|\boldsymbol{v}\|_{\left[k-N,k-1\right]}\right)^{p/a}g\left(k,l\right)^{p/a}}{N^q\,\underline{c}_p^{p/a}} +\\
        & \displaystyle \frac{c_{\beta}10^p\,5^{p/a}N^{p/a}\overline{\gamma}_{\alpha}\left(\|\boldsymbol{w_{\alpha}}\|_{\left[k-N,k-1\right]}\right)^{p/a}g\left(k,l\right)^{p/a}}{N^q\,\underline{c}_p^{p/a}}
    \end{array}
\end{equation}

Once we have found an upper bound for the first term of \eqref{eq:estimation error l-th it.}, we will follow a similar procedure to find a bound for the second and third terms. Using \eqref{eq:detectability} for the $l-th$ iteration, we can write
\begin{equation}
  \begin{array}{rl}
    \norm{\hat{w}_{j\vert k}} \stackrel{\eqref{eq:detectability}\eqref{property_1}}{\leq}& \underline{\gamma}_w^{-1}\left( \frac{5\overline{c}_p \norm{x_{k-N}-\bar{x}_{k-N}}^a g\left(k,l\right)}{N}\right)   \,   + \underline{\gamma}_w^{-1}\left( \frac{5\overline{c}_{\Lambda}\norm{\alpha_{k-N}-\bar{x}_{k-N}}^a g\left(k,l\right)}{N}\right) + \\
    & \underline{\gamma}_w^{-1}\left(\frac{5g\left(k,l\right)\overline{\gamma}_w\left(\|\boldsymbol{w}\|_{\left[k-N,k\right]}\right)}{N}\right)+\underline{\gamma}_w^{-1}\left(\frac{5g\left(k,l\right)\overline{\gamma}_v\left(\|\boldsymbol{v}\|_{\left[k-N,k\right]}\right)}{N} \, \right) + \\
    & \underline{\gamma}_{w}^{-1}\left(\frac{5g\left(k,l\right)\overline{\gamma}_{\alpha}\left(\|\boldsymbol{w_{\alpha}}\|_{\left[k-N,k\right]}\right)}{N}\right)
  \end{array}
\end{equation}
Introducing this bound in the second term of Equation \eqref{eq:estimation error l-th it.}:
\begin{align}
  \nonumber
  \gamma_1\left(\|\boldsymbol{w}-\hat{\boldsymbol{w}}\|_{\left[k-N,k\right]}\right) \leq& \gamma_1\left(\|\boldsymbol{w}\|_{\left[k-N,k\right]} + \|\hat{\boldsymbol{w}}\|_{\left[k-N,k\right]}\right) \\
  \nonumber
  \leq & \gamma_1\left(\|\boldsymbol{w}\|_{\left[k-N,k\right]} + \underline{\gamma}_w^{-1}\left( \frac{5\overline{c}_p \norm{x_{k-N}-\bar{x}_{k-N}}^a g\left(k,l\right)}{N}\right)+ \right.\\ \nonumber
  &\left. \underline{\gamma}_w^{-1}\left( \frac{5\overline{c}_{\Lambda}\norm{\alpha_{k-N}-\bar{x}_{k-N}}^a g\left(k,l\right)}{N}\right) + \right.\\ 
  &\left. \underline{\gamma}_w^{-1}\left(\frac{5g\left(k,l\right)\overline{\gamma}_w\left(\|\boldsymbol{w}\|_{\left[k-N,k\right]}\right)}{N}\right) + \right.\\ \nonumber
  &\left. \underline{\gamma}_w^{-1}\left(\frac{5g\left(k,l\right)\overline{\gamma}_v\left(\|\boldsymbol{v}\|_{\left[k-N,k\right]}\right)}{N}\right) + \right.\\ \nonumber
  &\left. \underline{\gamma}_{w}^{-1}\left(\frac{5g\left(k,l\right)\overline{\gamma}_{\alpha}\left(\|\boldsymbol{w_{\alpha}}\|_{\left[k-N,k\right]}\right)}{N}\right)
  \right)
\end{align}
Recalling Inequalities \eqref{property_1} we obtain the bound 
\begin{equation}        \label{eq:bound of gamma_1}
    \begin{array}{rl}
         \gamma_1\left( \|\boldsymbol{w}_j - \hat{\boldsymbol{w}}_{j\vert k}\|_{\left[k-N,k\right]} \right) \leq&  \gamma_1\left(6\|\boldsymbol{w}\|_{\left[k-N,k\right]}\right) + \gamma_1\left(6 \underline{\gamma}_w^{-1}\left( \frac{5\overline{c}_p \norm{x_{k-N}-\bar{x}_{k-N}}^a g\left(k,l\right)}{N}\right)\right) +\\
         & \gamma_1\left(6 \underline{\gamma}_w^{-1}\left( \frac{5\overline{c}_{\Lambda}\norm{\alpha_{k-N}-\bar{x}_{k-N}}^a g\left(k,l\right)}{N}\right) \right)+\\
         & \gamma_1\left(6 \underline{\gamma}_w^{-1}\left(\frac{5g\left(k,l\right)\overline{\gamma}_w\left(\|\boldsymbol{w}\|_{\left[k-N,k\right]}\right)}{N}\right) \right) + \\
         & \gamma_1\left(6 \underline{\gamma}_w^{-1}\left(\frac{5g\left(k,l\right)\overline{\gamma}_v\left(\|\boldsymbol{v}\|_{\left[k-N,k\right]}\right)}{N}\right) \right) + \\
         & \gamma_1\left(6 \underline{\gamma}_{w}^{-1}\left(\frac{5g\left(k,l\right)\overline{\gamma}_{\alpha}\left(\|\boldsymbol{w_{\alpha}}\|_{\left[k-N,k\right]}\right)}{N}\right) \right).
    \end{array}
\end{equation}
With the use of Assumption \ref{stage cost assumption}, the bound can be finally written as
\begin{equation}        \label{eq:final bound of gamma_1}
    \begin{array}{rl}
         \gamma_1\left( \|\boldsymbol{w}_j - \hat{\boldsymbol{w}}_{j\vert k}\|_{\left[k-N,k\right]} \right) \leq&  \gamma_1\left(6\|\boldsymbol{w}\|_{\left[k-N,k\right]}\right)+ \displaystyle \frac{c_1 5^{b_1}\overline{c}_p^{b_1}\norm{x_{k-N}-\bar{x}_{k-N}}^{ab_1}g\left(k,l\right)^{b_1}}{N^{b_1}}+\\
         & \displaystyle \frac{c_1 5^{b_1}\overline{c}_{\Lambda}^{b_1}\norm{\alpha_{k-N}-\bar{\alpha}_{k-N}}^{ab_1}g\left(k,l\right)^{b_1}}{N^{b_1}}+ \\
         & \displaystyle c_1 5^{b_1}g\left(k,l\right)^{b_1}\overline{\gamma}_w\left(\|\boldsymbol{w}\|_{\left[k-N,k\right]}\right)^{b_1}+\\
         & \displaystyle c_1 5^{b_1}g\left(k,l\right)^{b_1}\overline{\gamma}_v\left(\|\boldsymbol{v}\|_{\left[k-N,k\right]}\right)^{b_1}+\\
         & \displaystyle c_1 5^{b_1}g\left(k,l\right)^{b_1}\overline{\gamma}_{\alpha}\left(\|\boldsymbol{w_{\alpha}}\|_{\left[k-N,k\right]}\right)^{b_1}
    \end{array}
\end{equation}
With a similar procedure, a bound for the third term of inequality \eqref{eq:estimation error l-th it.} is found
\begin{equation}        \label{eq:final bound of gamma_2}
    \begin{array}{rl}
         \gamma_2\left( \|\boldsymbol{v}_j - \hat{\boldsymbol{v}}_{j\vert k}\|_{\left[k-N,k\right]} \right) \leq&  \gamma_2\left(6\|\boldsymbol{v}\|_{\left[k-N,k\right]}\right)+ \displaystyle \frac{c_2 5^{b_2}\overline{c}_p^{b_2}\norm{x_{k-N}-\bar{x}_{k-N}}^{ab_2}g\left(k,l\right)^{b_2}}{N^{b_2}}+\\
         & \displaystyle \frac{c_2 5^{b_2}\overline{c}_{\Lambda}^{b_2}\norm{\alpha_{k-N}-\bar{\alpha}_{k-N}}^{ab_2}g\left(k,l\right)^{b_2}}{N^{b_2}}+ \\
         & \displaystyle c_2 5^{b_2}g\left(k,l\right)^{b_2}\overline{\gamma}_w\left(\|\boldsymbol{w}\|_{\left[k-N,k\right]}\right)^{b_2}+\\
         & \displaystyle c_2 5^{b_2}g\left(k,l\right)^{b_2}\overline{\gamma}_v\left(\|\boldsymbol{v}\|_{\left[k-N,k\right]}\right)^{b_2}+\\
         & \displaystyle c_2 5^{b_2}g\left(k,l\right)^{b_2}\overline{\gamma}_{\alpha}\left(\|\boldsymbol{w_{\alpha}}\|_{\left[k-N,k\right]}\right)^{b_2}
    \end{array}
\end{equation}
The estimation error given in Equation \eqref{eq:estimation error l-th iteration} can be bounded as
\begin{equation}    \label{eq:estimation error l-th iteration}
  \begin{array}{rl}
    \norm{x_k-\hat{x}_{k\vert k}} \leq & \displaystyle \frac{\norm{x_{k-N}-\bar{x}_{k-N}}^p}{N^q}\left(
        \displaystyle c_{\beta}2^p + \frac{c_{\beta}10^p\,5^{p/a}\overline{c}_p^{p/a}g\left(k,l\right)^{p/a}}{\underline{c}_p^{p/a}}\right)  +\\
        & \displaystyle \frac{c_{\beta}10^p\,5^{p/a}\overline{c}_{\Lambda}^{p/a}\norm{\alpha_{k-N}-\bar{\alpha}_{k-N}}^p g\left(k,l\right)^{p/a}}{N^q\,\underline{c}_p^{p/a}} +\\
        & \displaystyle \frac{c_{\beta}10^p\,5^{p/a}N^{1/a}\overline{\gamma}_w\left(\|\boldsymbol{w}\|_{\left[k-N,k-1\right]}\right)^{1/a}g\left(k,l\right)^{1/a}}{N^q\,\underline{c}_p^{p/a}} +\\
        & \displaystyle \frac{c_{\beta}10^p\,5^{p/a}N^{p/a}\overline{\gamma}_v\left(\|\boldsymbol{v}\|_{\left[k-N,k-1\right]}\right)^{p/a}g\left(k,l\right)^{p/a}}{N^q\,\underline{c}_p^{p/a}} +\\
        & \displaystyle \frac{c_{\beta}10^p\,5^{p/a}N^{p/a}\overline{\gamma}_{\alpha}\left(\|\boldsymbol{w_{\alpha}}\|_{\left[k-N,k-1\right]}\right)^{p/a}g\left(k,l\right)^{p/a}}{N^q\,\underline{c}_p^{p/a}}+\\
        & \gamma_1\left(6\|\boldsymbol{w}\|_{\left[k-N,k\right]}\right)+ \displaystyle \frac{c_1 5^{b_1}\overline{c}_p^{b_1}\norm{x_{k-N}-\bar{x}_{k-N}}^{ab_1}g\left(k,l\right)^{b_1}}{N^{b_1}}+\\
         & \displaystyle \frac{c_1 5^{b_1}\overline{c}_{\Lambda}^{b_1}\norm{\alpha_{k-N}-\bar{\alpha}_{k-N}}^{ab_1}g\left(k,l\right)^{b_1}}{N^{b_1}}+ \\
         & \displaystyle c_1 5^{b_1}g\left(k,l\right)^{b_1}\overline{\gamma}_w\left(\|\boldsymbol{w}\|_{\left[k-N,k\right]}\right)^{b_1}+\\
         & \displaystyle c_1 5^{b_1}g\left(k,l\right)^{b_1}\overline{\gamma}_v\left(\|\boldsymbol{v}\|_{\left[k-N,k\right]}\right)^{b_1}+\\
         & \displaystyle c_1 5^{b_1}g\left(k,l\right)^{b_1}\overline{\gamma}_{\alpha}\left(\|\boldsymbol{w_{\alpha}}\|_{\left[k-N,k\right]}\right)^{b_1} + \\
         & \gamma_2\left(6\|\boldsymbol{v}\|_{\left[k-N,k\right]}\right)+ \displaystyle \frac{c_2 5^{b_2}\overline{c}_p^{b_2}\norm{x_{k-N}-\bar{x}_{k-N}}^{ab_2}g\left(k,l\right)^{b_2}}{N^{b_2}}+\\
         & \displaystyle \frac{c_2 5^{b_2}\overline{c}_{\Lambda}^{b_2}\norm{\alpha_{k-N}-\bar{\alpha}_{k-N}}^{ab_2}g\left(k,l\right)^{b_2}}{N^{b_2}}+ \\
         & \displaystyle c_2 5^{b_2}g\left(k,l\right)^{b_2}\overline{\gamma}_w\left(\|\boldsymbol{w}\|_{\left[k-N,k\right]}\right)^{b_2}+\\
         & \displaystyle c_2 5^{b_2}g\left(k,l\right)^{b_2}\overline{\gamma}_v\left(\|\boldsymbol{v}\|_{\left[k-N,k\right]}\right)^{b_2}+\\
         & \displaystyle c_2 5^{b_2}g\left(k,l\right)^{b_2}\overline{\gamma}_{\alpha}\left(\|\boldsymbol{w_{\alpha}}\|_{\left[k-N,k\right]}\right)^{b_2}
    
    \end{array}
\end{equation}
Since the vector $\alpha$ (and its estimated $\hat{\alpha}$) satisfies $\sum_{i=1}^{q}\alpha_i=1$ and $\alpha_i\geq 0$, the maximal value of $\norm{\alpha-\hat{\alpha}}$ is upper bounded by $\sqrt{2}$, i.e., $\max\{\norm{\alpha-\hat{\alpha}}\} = \max\{\norm{w_{\alpha}}\}\leq \sqrt{2}$.
Defining the constants as follows
\begin{equation*}
  q\geq p/a, \; \zeta \coloneqq \max \left \{p, a\,b_1, a\,b_2\right \} \; \eta \coloneqq \min\left\{ q, b_1, b_2\right\},    
\end{equation*}
inequality \eqref{eq:estimation error l-th iteration} can be rewritten as follows 
\begin{equation}    \label{eq:estimation error l-th iteration conservative bound}
  \begin{array}{rl}
    \norm{x_k - \hat{x}_{k\vert k}} \leq& \displaystyle\frac{\norm{x_{k-N}-\bar{x}_{k-N}}^{\zeta}}{N^{\eta}} \left( \left(1 + \frac{5^{p+q}\overline{c}_p^{q} g\left(k,l\right)^q}{\underline{c}_p^{p/a}}\right)c_{\beta}2^p + c_1 5^{b_1}\overline{c}_1^{b_1}g\left(k,l\right)^{b_1} + \right.\\
    & \left. c_2 5^{b_2}\overline{c}_p^{b_2}g\left(k,l\right)^{b_2} \right) + \displaystyle\frac{\norm{\alpha_{k-N}-\bar{\alpha}_{k-N}}^{\zeta}}{N^{\eta}}\left( \frac{c_{\beta}10^p5^q\overline{c}_{\Lambda}^qg\left(k,l\right)^q}{\overline{c}_p^{p/a}} + \right.\\
    & \left. c_1 5^{b_1}\overline{\gamma}_{\alpha}\left(\sqrt{2}\right)^{b_1} + c_25^{b_2}\overline{\gamma}_{\alpha}\left(\sqrt{2}\right)^{b_2} \right) + g\left(k,l\right)^{\eta}\left( \frac{c_{\beta}10^p5^q\overline{\gamma}_{\alpha}\left(\sqrt{2}\right)^q}{\underline{c}_p^{p/a}} + \right. \\
    & \left. c_15^{b_1}\overline{\gamma}_{\alpha}\left(\sqrt{2}\right)^{b_1} + c_25^{b_2}\overline{\gamma}_{\alpha}\left(\sqrt{2}\right)^{b_2} \right) + \frac{c_{\beta}10^p5^qg\left(k,l\right)^q\overline{\gamma}_w\left(\|\boldsymbol{w}\|_{\left[k-N,k\right]}\right)^q}{\underline{c}_p^{p/a}} + \\
    & \gamma_1\left(6\|\boldsymbol{w}\|_{\left[k-N,k\right]}\right) + c_15^{b_1}g\left(k,l\right)^{b_1}\overline{\gamma}_w\left(\|\boldsymbol{w}\|_{\left[k-N,k\right]}\right)^{b_1}+\\
    & c_25^{b_2}g\left(k,l\right)^{b_2}\overline{\gamma}_w\left(\|\boldsymbol{w}\|_{\left[k-N,k\right]}\right)^{b_2} + \frac{c_{\beta}10^p5^q g\left(k,l\right)^q\overline{\gamma}_v\left(\|\boldsymbol{v}\|_{\left[k-N,k\right]}\right)^q}{\underline{c}_p^{p/a}} + \\
    & \gamma_2\left(6\|\boldsymbol{v}\|_{\left[k-N,k\right]}\right) + c_15^{b_1}g\left(k,l\right)^{b_1}\overline{\gamma}_v\left(\|\boldsymbol{v}\|_{\left[k-N,k\right]}\right)^{b_1} + \\
    & c_25^{b_2}g\left(k,l\right)^{b_2}\overline{\gamma}_v\left(\|\boldsymbol{v}\|_{\left[k-N,k\right]}\right)^{b_2}.
    \end{array}
\end{equation}
Noting again that $\norm{\alpha_{k-N}-\bar{\alpha}_{k-N}}\leq \sqrt{2}$ and defining the functions and constants as follows
\begin{align}
    k_{1} \coloneqq & c_{\beta}2^p \\ 
    k_{2}\coloneqq & \left(5^{p+q}\left(\frac{\overline{c}_p}{\underline{c}_p}\right)^qc_{\beta}2^p+   c_15^{b_1}\overline{c}_p^{b_1}+c_25^{b_2}\overline{c}_p^{b_2}\right), \\
    K \coloneqq & \frac{c_{\beta}10^p5^q}{\underline{c}_p^{p/a}}+c_15^{b_1}+c_25^{b_2} \\
%
    \psi_{w1} \coloneqq & \gamma_1\left(6\|\boldsymbol{w}\|_{\left[k-N,k\right]}\right), \\
    \psi_{w2} \coloneqq & \frac{c_{\beta}10^p5^q\overline{\gamma}_w\left(\|\boldsymbol{w}\|_{\left[k-N,k\right]}\right)^q}{\underline{c}_p^{p/a}} + \left(c_15^{b_1} + c_25^{b_2}\right) \overline{\gamma}_w\left(\|\boldsymbol{w}\|_{\left[k-N,k\right]}\right)^{b_1} \\
    \psi_{v1} \coloneqq & \gamma_2\left(6\|\boldsymbol{v}\|_{\left[k-N,k\right]}\right), \\
    \psi_{w2} \coloneqq & \frac{c_{\beta}10^p5^q\overline{\gamma}_v\left(\|\boldsymbol{v}\|_{\left[k-N,k\right]}\right)^q}{\underline{c}_p^{p/a}} + \left(c_15^{b_1} + c_25^{b_2}\right) \overline{\gamma}_v\left(\|\boldsymbol{v}\|_{\left[k-N,k\right]}\right)^{b_1} \\
\end{align}
the bound of the estimation error can be rewritten as follows
\begin{equation} \label{eq:bound of estimation error pre final}
  \begin{array}{rl}
    \norm{x_k -\hat{x}_{k\vert k}} \leq & \displaystyle\frac{\norm{x_{k-N}-\bar{x}_{k-N}}^{\zeta}}{N^{\eta}}\left(k_1 + g\left(k,l\right)^{\eta}k_2\right) + \psi_{w1}+g\left(k,l\right)\psi_{w2}+ \\
    & \psi_{v1}+g\left(k,l\right)\psi_{v2}+g\left(k,l\right)\left(\overline{\gamma}_{\alpha}\left(\sqrt{2}\right)^{\zeta} + \displaystyle\frac{\overline{c}_{\Lambda}2^{\zeta/2}}{N^{\eta}}\right)K
  \end{array}
\end{equation}
Defining others constants and functions
\begin{align}
    \bar{k}_{\beta}\left(l\right) \coloneqq & k_1 + g\left(k,l\right)k_2      \\
    \Phi_w\left(w,l\right) \coloneqq & \psi_{w1} + g\left(k,l\right)^{\eta}\psi_{w2} \\
    \Phi_v\left(v,l\right) \coloneqq & \psi_{v1} + g\left(k,l\right)^{\eta}\psi_{v2} \\
    \Phi_{\alpha}\left(l,N\right) \coloneqq & g\left(k,l\right)\left(\overline{\gamma}_{\alpha}\left(\sqrt{2}\right)^{\zeta} + \frac{\overline{c}_{\Lambda}^{\zeta}\,2^{\zeta/2}}{N^{\eta}}\right) K \\
    \bar{\beta}\left(r,s\right) \coloneqq & \frac{\bar{k}_{\beta}r^p}{s^q}
\end{align}
with $p=\zeta$, $q=\eta$, $\bar{\beta}\left(r,s\right)\in \mathcal{KL}$, one can write the estimation error as follows
\begin{equation}
\label{eq:RAS}
  \begin{array}{rl}
    \norm{x_k -\hat{x}_{k\vert k}} \leq & \bar{\beta}\left(\norm{x_{k-N}-\bar{x}_{k-N}}, N\right) + \Phi_w\left(\|\boldsymbol{w}\|_{\left[k-N,k\right]},l\right) + \Phi_v\left(\|\boldsymbol{v}\|_{\left[k-N,k\right]},l\right) +\\
    & \Phi_{\alpha}\left(l, N\right)
  \end{array}
\end{equation}
The reader can verify that the same result is obtained for $k\in \mathbb{Z}_{\left[1,N-1\right]}$. To guarantee the validity of previous results to the entire time horizon the definition of $\beta\left( r, s\right)$  must be extended to $s = 0$ . Because of $\bar{\beta}\left(r, s \right) \in \mathcal{KL}$, $\bar{\beta}\left(r, 0 \right) \in \mathcal{KL}$ and $\bar{\beta}\left(r, 0 \right) \geq \bar{\beta}\left(r, k \right)$, $\forall k \in \mathbb{Z}_{\geq 1}$, it is sufficient to define $\bar{\beta}\left(r, 0 \right) \coloneqq k_{\beta} \; \bar{\beta}\left(r, 1 \right)$ for some $k_{\beta} \in \mathbb{R}_{> 1}$ to extend the definition of these function for all $k \in \mathbb{Z}_{\left[0,N\right]}$.

Let us select some $\epsilon \in \mathbb{R}_{>0}$ and
\begin{equation*}
  \begin{array}{rl}
     r_{\max}\coloneqq& \left\{ \displaystyle \bar{\beta}\left(e_{\max},0\right)+\Phi_w\left(w_{\max},1\right)+\Phi_v\left(v_{\max},1\right)+\Phi_{\alpha}\left(1,\mathcal{N}\right),\right.\\
     & \left.\left(1+\epsilon\right)\left(\Phi_w\left(w_{\max},1\right)+\Phi_v\left(v_{\max},1\right)+\Phi_{\alpha}\left(1,\mathcal{N}\right)\right) \right\}
  \end{array}
\end{equation*}
Let us define $\mathcal{N}$ as
\begin{equation}
    \begin{array}{rl}
         \displaystyle\left(\frac{2\left(1+\epsilon\right)\left(k_1+k_2g\left(k,l\right)^{\eta}\right)e_{\max}^{\zeta}}{r_{\max}}\right)^{1/\eta} \leq & \mathcal{N}
    \end{array}
\end{equation}

\begin{remark}
Note that due to the model uncertainty, the horizon must be enlarged.
\end{remark}
Adopting an estimator with a window length greater or equal to $\mathcal{N}$, one will have
\begin{equation}		\label{beta equiv for N0}
  \bar{\beta}\left(r, \; N \right) \leq \frac{r}{2},
\end{equation}
the effects of the initial conditions will vanish. As $k \rightarrow \infty$, the estimation error will entry to the bounded set $\mathcal{X}\left(w,v \right) \subset \mathcal{X}$ defined by the noises of the system and the uncertainty
\begin{equation}
  \mathcal{X} \left(w,v,l \right) \coloneqq \{ \norm{x_{k} - 
  \hat{x}_{k\vert k}} \leq \left(1 + \epsilon \right)\left(\Phi_w \left(\supofseq{w},1 \right) + \right.
  \left. \Phi_v\left(\supofseq{v},1 \right) +\Phi_{\alpha}\left(1,\mathcal{N}\right)\right) \}.
\end{equation}
This set define the minimum size region of error space $\mathcal{X}$ that the error can achieve by removing the effect of errors in initial conditions ($e_{max}$). Equation \eqref{beta equiv for N0} establish a trade off between speed of convergence and window length, which is related with the size of $\mathcal{X}\left(w,v,l \right)$.

For any MHE with adaptive arrival cost and window length $N \geq \mathcal{N}$ two situations can be considered
\begin{itemize}
  \item The estimator has removed the effects of $x_0$ on $\hat{x}_{k\vert k}$ such that
  $\norm{x_{k}-\hat{x}_{k\vert k}} \in \mathcal{X}\left(w,v,l \right)$, and
  \item The estimator has not removed the effects of $x_0$ on $\hat{x}_{k \vert k}$ such that $\norm{x_{k}-\hat{x}_{k\vert k}}
  \notin \mathcal{X}\left(w,v,l \right)$,
\end{itemize}

Let us assume that the estimation error is
\begin{equation*}
  \norm{x_k-\hat{x}_{k\vert k}}\leq 2\left(1+\epsilon\right)\left(\Phi_w\left(w_{\max},1\right)+\Phi_v\left(v_{\max},1\right)+\Phi_{\alpha}\left(1,\mathcal{N}\right)\right),    
\end{equation*}
the estimation error will be given by
\begin{align}
   \norm{x_{k+N}-\hat{x}_{k+N\vert k+N}} \leq&\, \bar{\beta}\left(\norm{x_k-\hat{x}_{k\vert k+N}}, \mathcal{N}\right)
   + \Phi_w\left(w_{\max},1\right)+\nonumber\\
   & \Phi_v\left(v_{\max},1\right)+\Phi_{\alpha}\left(1,\mathcal{N}\right),\nonumber \\
   \leq &\, \frac{\norm{x_k-\hat{x}_{k\vert k}}}{2\left(1+\epsilon\right)} +\Phi_w\left(w_{\max},1\right)+ \Phi_v\left(v_{\max},1\right)+\Phi_{\alpha}\left(1,\mathcal{N}\right)\nonumber\\
   \leq &\, 2\left(\Phi_w\left(w_{\max},1\right)+ \Phi_v\left(v_{\max},1\right)+\Phi_{\alpha}\left(1,\mathcal{N}\right)\right)\nonumber \\
   \leq &\, 2\left(1+\epsilon\right)\left(\Phi_w\left(w_{\max},1\right)+ \Phi_v\left(v_{\max},1\right)+\Phi_{\alpha}\left(1,\mathcal{N}\right)\right)
\end{align}
Therefore, when
\begin{equation*}
  \norm{x_k-\hat{x}_{k\vert k}}\leq 2\left(1+\epsilon\right)\left(\Phi_w\left(w_{\max},1\right)+\Phi_v\left(v_{\max},1\right)+\overline{K}\left(1\right)\right),
\end{equation*}
the error will no become larger. Assuming now
\begin{equation*}
  r_{\max}\geq\norm{x_k -\hat{x}_{k\vert k}}>2\left( 1 + \epsilon \right) \left(\Phi_w\left(w_{\max}\right)+\Phi_v\left(v_{\max}\right)+\Phi_{\alpha}\left(1,\mathcal{N}\right)\right)
\end{equation*}
the estimation error is given by
\begin{equation}
  \begin{array}{rl}
    \norm{x_{k+N}-\hat{x}_{k+N\vert k+N}} \leq& \bar{\beta}\left(\norm{x_k-\hat{x}_{k\vert k+N}}, \mathcal{N}\right)+\Phi_w\left(w_{\max},1\right)+\nonumber\\
    & \Phi_v\left(v_{\max},1\right)+\Phi_{\alpha}\left(1,\mathcal{N}\right)\nonumber\\
    \leq& \frac{\norm{x_k-\hat{x}_{k\vert k+N}}}{2\left(1+\epsilon\right)} +  \Phi_w\left(w_{\max},1\right)+\Phi_v\left(v_{\max},1\right)+\Phi_{\alpha}\left(1,\mathcal{N}\right)\\
    \leq& \frac{\norm{x_k-\hat{x}_{k\vert k+N}}}{2\left(1+\epsilon\right)}+ \frac{\norm{x_k-\hat{x}_{k\vert k+N}}}{2\left(1+\epsilon\right)}\nonumber\\
    \leq & \norm{x_k-\hat{x}_{k\vert k+N}}\left(\frac{1}{1+\epsilon}\right)
  \end{array}
\end{equation}
with $\xi\coloneqq\left(\frac{1}{1+\epsilon}\right)<1$, since $\epsilon>0$

In the latter case, the estimator error behaves contractively. By mean of some definitions, i.e., $i\coloneqq \lfloor \frac{k}{\mathcal{N}} \rfloor$, $j\coloneqq t \mod \mathcal{N}$, time $k$ can be expressed as $k=i\mathcal{N}+j$. For $N\geq\mathcal{N}$, the $\mathcal{KL}$ functions $\bar{\beta}\left(r,s\right)$ is decreasing every $\mathcal{N}$ samples. Writing the estimation error with this notation for time $k$
\begin{equation}
  \begin{array}{rl}
    \norm{x_k-\hat{x}_{k\vert k}} \leq& \max\left\{ \xi^i\norm{x_j-\hat{x}_{j\vert k}}, 2\left(1+\epsilon\right) \left(\Phi_w\left(w_{\max},1\right)+\Phi_v\left(v_{\max},1\right)+\Phi_{\alpha}\left(1,\mathcal{N}\right)\right) \right\} \\
    \leq& \bar{\beta}\left(\norm{x_0-\bar{x}_0},j\right)\xi^i  + 2\left(1+\mu\right) \left(\Phi_w\left(w_{\max},1\right)+\Phi_v\left(v_{\max},1\right)+\Phi_{\alpha}\left(1,\mathcal{N}\right)\right) \\
  \end{array}
\end{equation}
Defining
\begin{equation}
   \begin{array}{rl}
      \overline{\Phi}\left(\norm{x_0-\bar{x}_0}\right) \coloneqq& \xi^i\bar{\beta}\left(\norm{x_0-\bar{x}_0},j\right), \\
      \overline{\Phi}_w\left(w_{\max}\right) \coloneqq& 2\left(1+\mu\right)\Phi_w\left(w_{\max},1\right), \\
      \overline{\Phi}_v\left(v_{\max}\right) \coloneqq& 2\left(1+\mu\right)\Phi_v\left(v_{\max},1\right), \\
      \overline{\Phi}_{\alpha}\left(l_{\min},\mathcal{N}\right)  \coloneqq& 2\left(1+\mu\right)\Phi_{\alpha}\left(1,\mathcal{N}\right).
  \end{array}
\end{equation}
Finally we can write
\begin{equation}
  \norm{x_k-\hat{x}_{k\vert k}} \leq  \overline{\Phi}\left(\norm{x_0-\bar{x}_0},k\right) + \overline{\Phi}_w\left(w_{\max}\right) + \overline{\Phi}_v\left(v_{\max}\right) + \overline{\Phi}_{\alpha}\left(l_{\min},\mathcal{N}\right)
\end{equation}
Taking $w_{\max}$ from $\|\boldsymbol{w}\|_{\left[0,k\right]}$ instead $\|\boldsymbol{w}\|_{\left[k-N,k\right]}$ and $v_{\max}$ from $\|\boldsymbol{v}\|_{\left[0,k\right]}$ instead $\|\boldsymbol{v}\|_{\left[k-N,k\right]}$, and noting that Equation \eqref{eq:RAS} still being valid, the robust regional practical stability is proved.

On the other hand, the convergence of the estimator to the true state in the case of decaying disturbances can be established.
Assuming $\lim_{k\rightarrow\infty}w_k=0$ and $\lim_{k\rightarrow\infty}v_k=0$, given Equation \eqref{eq:RAS}, one can choose some $k\geq K_1$ for which $\max\left\{ w_{\max},v_{\max} \right\} \leq \min \left\{ \Phi_w^{-1}\left(\frac{\varepsilon}{4}\right), \Phi_v^{-1}\left(\frac{\varepsilon}{4}\right) \right\}$. At the same time, one can choose some large enough value of $l$ such that $\overline{\Phi}_{\alpha}\left(l,\mathcal{N}\right) \leq \frac{\varepsilon}{4}$. Note that according to Equation \eqref{g(k,l)}, the value of $l$ will be getting smaller as $\|\boldsymbol{w}\|\rightarrow 0$ and $\|\boldsymbol{v}\|\rightarrow 0$.
Recalling that $\overline{\Phi}\left(\cdot\right) \in \mathcal{KL}$, there will exist  some $k\geq K_2$ such that $\overline{\Phi}\left(\norm{x_{k-K_2}-\bar{x}_{k-K_2}},k\right) \leq \frac{\varepsilon}{4}$. Under these conditions, there exists some time $k\geq \max\left\{K_1,K_2 \right\}+\mathcal{N}$ such that
\begin{equation}
    \begin{array}{rl}
         \norm{x_k-\hat{x}_{k\vert k}} \leq & \overline{\Phi}\left(\norm{x_{k-\mathcal{N}}-\bar{x}_{k-\mathcal{N}}},k\right) + \overline{\Phi}_w\left(w_{\max}\right) + \overline{\Phi}_v\left(v_{\max}\right) + \overline{\Phi}_{\alpha}\left(l,\mathcal{N}\right) \\
         \norm{x_k-\hat{x}_{k\vert k}} \leq & \displaystyle\frac{\varepsilon}{4}+\frac{\varepsilon}{4}+\frac{\varepsilon}{4}+\frac{\varepsilon}{4} \\
         \norm{x_k-\hat{x}_{k\vert k}} \leq & \varepsilon
    \end{array}
\end{equation}
Since one can choose any value of $\displaystyle\varepsilon$, $\lim_{k\rightarrow \infty} \norm{x_k-\hat{x}_{k\vert k}} = 0$ can be guaranteed when $\lim_{k\rightarrow\infty}w_k=0$ and $\lim_{k\rightarrow\infty}v_k=0$ as claimed.

\qed

\begin{remark}
As expected, the model uncertainty deteriorates state estimation. However, for a large enough value of $l$, this effect can be mitigated. Moreover, when $\lim_{l\rightarrow\infty}\Phi_{\alpha}\left(l,\mathcal{N}\right) = 0$
\end{remark}

\section{Simulation and results}
The following examples will be used to illustrate the results presented in the previous sections and to evaluate and compare the performance of the proposed estimator with others from the state of the art. The process disturbance is white Gaussian noise acting as an additive exogenous input to the system.

\subsection{Unknown linear system}
Let us considers the linear system 
\begin{align*}
	x_{k+1} &= A_p x_k + w_k \\
    y_k 	&= C_p x_k + v_k
\end{align*}
whose matrices are unknown an they have the following structure
\begin{align}       \label{linear sys}
    A_p &= \left[ \begin{array}{cc}
         0 	& 	a_1  \\
         1 	& 	a_2
    \end{array} \right],
    C_p = \left[ \begin{array}{cc}
    c_1 & c_2
    \end{array} \right]
\end{align}
The system is affected with additive process and measurement noise $w$ and $v$ drawn from normal distributions with zero mean and covariance $Q_w = S_w^2 I_2$ and $R_v = S_v^2$, respectively. The polytope is defined using three \emph{LTI} models
\begin{equation}        \label{linear_pol}
  \begin{array}{c}
    A_1 = \left[ \begin{array}{cc}
         0 	& 	0.72  \\
         1 	& 	0.28
    \end{array} \right],
    A_2 = \left[ \begin{array}{cc}
         0 	&  -0.59  \\
         1 	& 	1.57
    \end{array} \right],
    A_3 = \left[ \begin{array}{cc}
         0 	& 	-0.35  \\
         1 	& 	1.26
    \end{array} \right], \\
    C_1 = \left[ \begin{array}{cc}
    -1.46 & -1.29 \\
    \end{array} \right],
    C_2 = \left[ \begin{array}{cc}
    -4.84 & -2.90 \\
    \end{array} \right],
    C_3 = \left[ \begin{array}{cc}
    -0.09 & -0.03
    \end{array} \right],
  \end{array}
\end{equation}
such that the system belongs to it. 
The matrices of the system $\left(A_{S}, C_{S} \right)$ and its model $\left(A_{M}, C_{M} \right)$ were generated as a convex combination of the polytope with mixing parameters $\mathbf{\alpha_{S}}$ ($\alpha_{S,1}=0.22, \alpha_{S,2}=0.76, \alpha_{S,3}=0.02$) and $\mathbf{\alpha_{M}}$ ($\alpha_{M,1}=0.41, \alpha_{M,2}=0.22, \alpha_{M,3}=0.37$), respectively.

%
The stage cost of the receding horizon estimators is chosen as $\ell(w,v) = w^T Q^{-1} w + v^T R^{-1} v$ with $R^{-1} = 5$ and $Q^{-1}=diag\left(0.1, 0.1\right)$. 
The proposed moving horizon estimator ($MHE_{A}$) the prior weighting matrix is given by $\Gamma_{k-N} \left(\chi\right) = (\chi - \hat{x}(0|k))^T P^{-1}_{k|k} (\chi - \hat{x}(0|k))$, where $P^{-1}_0 = 0.1 \boldsymbol{I}_2$ and $P_{k|k}$ is updated using equations \eqref{eq:updatePk} and \eqref{eq:updateXA} with $\sigma = 1e^{-4}$ and $c=5$.
The robust moving horizon estimator ($MHE_{R}$) implements the algorithm proposed by \cite{muller2017nonlinear} with the nominal model ($A_M,C_M$), the prior weight given by $\Gamma \left(\chi\right) = L(\chi - \hat{x}(0|k))^T (\chi - \hat{x}(0|k))$ and parameters $\delta = 1$, $\delta_1 = \kappa^N$ ($\kappa=0.89$) and $\delta_2 = 1/N$ (see equation (3) of \cite{muller2017nonlinear}). 
The full information (\emph{FIE}, see \cite{ji2016robust}) is configured with the true model and the same parameters used by the $MHE_{R}$ with $\delta = 1$, $\delta_1 = \kappa^k$ and $\delta_2 = 1/k$ with the system matrices ($A_S,C_S$) 
The robust Kalman filter ($KF_R$) was designed following the design procedure proposed by \cite{zhu2002design} using the nominal model ($A_M,C_M$) and computing the bounds from the models of the polytope. The Kalman filter ($KF$) was designed using the matrices of the system ($A_S,C_S$)

\begin{table}[th]
  \centering  
  \caption{Averaged MSE for $S_w = 0.1, S_v = 0.05, N=8$.}
  \begin{tabular}{|l|c|c|c|c|c|}						\hline
	 & \small$KF_{R}$ & \small $KF$ &\small $FIE$ &\small $MHE_{R}$ &\small $MHE_{A}$        \\ 	\hline
     $x_0$ 	&\small0.78467 &\small0.030851 &\small0.0054 &\small0.2662 &\small0.0212			\\ 	\hline
     $x_1$ 	&\small1.9946  &\small0.069122 &\small0.0039 &\small0.4675 &\small0.0389 			\\ 	\hline
  \end{tabular}       \label{tab:linear system 1}
\end{table}
Table \ref{tab:linear system 1} shows the mean square estimation error ($MSE$) of each estimator averaged over 100 trials for $S_w = 0.1$, $S_v = 0.05$ and  $N=8$ for all receding horizon estimators ($FIE$, $MHE_R$ and $MHE_A$). It can be seen that the proposed estimator average mean square estimation error is smaller than the Kalman filters ($KF$ and $KF_R$) and $MHE_R$. Only the $FIE$ provides better performance than the proposed algorithm. The performance difference between the estimators that employ the nominal model ($KF_R$ and $MHE_R$) is due to the adaptation capabilities of $MHE_A$ that allows to reduce the uncertainty of the estimator model. The main performance difference between $MHE_A$ and $FIE$ estimators is due to the model employed by estimator and the amount of information employed to estimate $\hat{x}_{k|k}$. While the $FIE$ estimator use of the exact model and all the system output available until $k$, the $MHE_A$ identifies the model in the initial samples and only use the last $N$ system outputs.

\begin{figure}[tb]
  \centering
  \includegraphics[width=0.90\textwidth]{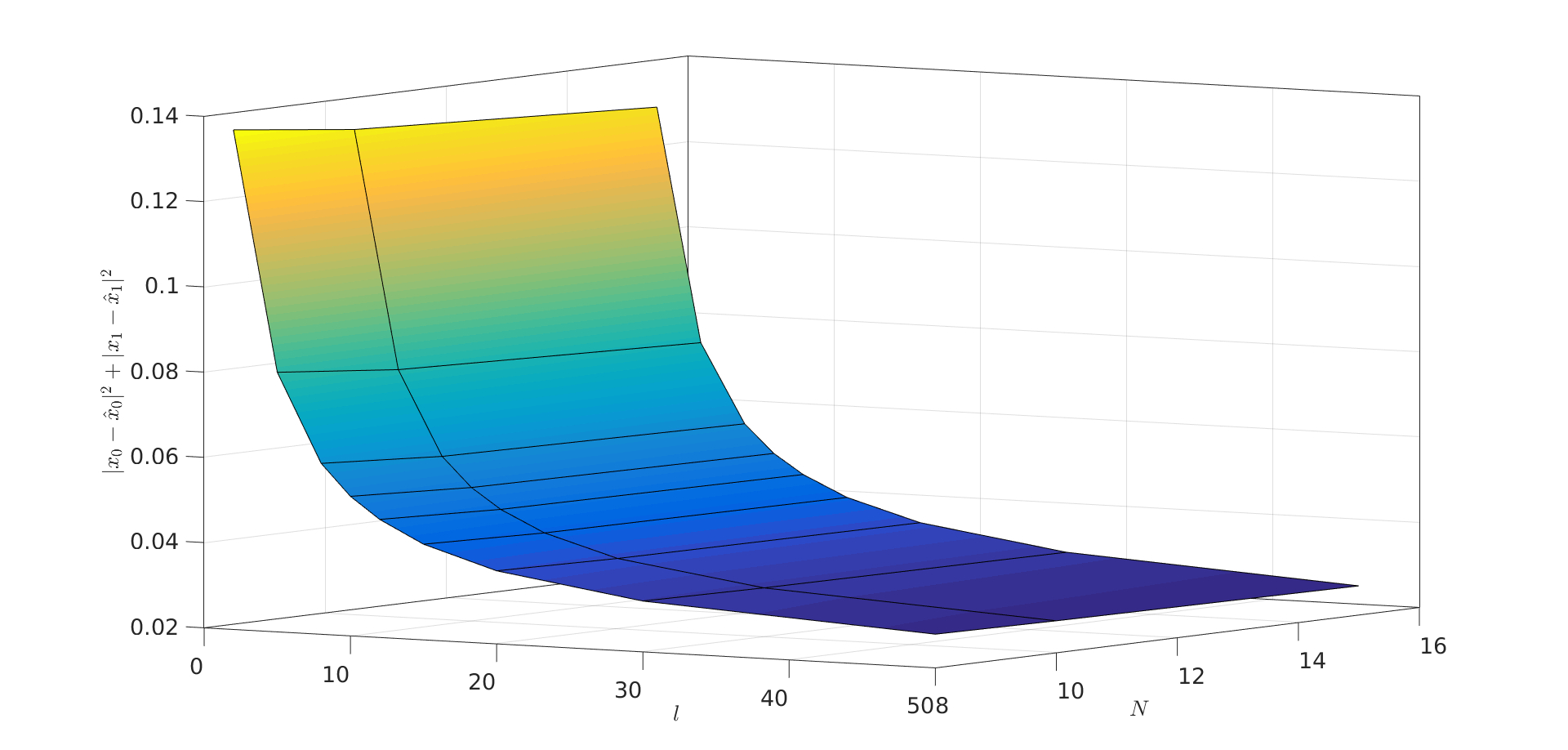}
  \caption{Estimation error for different values of $l$ and $N$.}
  \label{fig:decreasing cost with l 3D}
\end{figure}
\begin{figure}[b]
 \centering
 \includegraphics[width=0.80\textwidth]{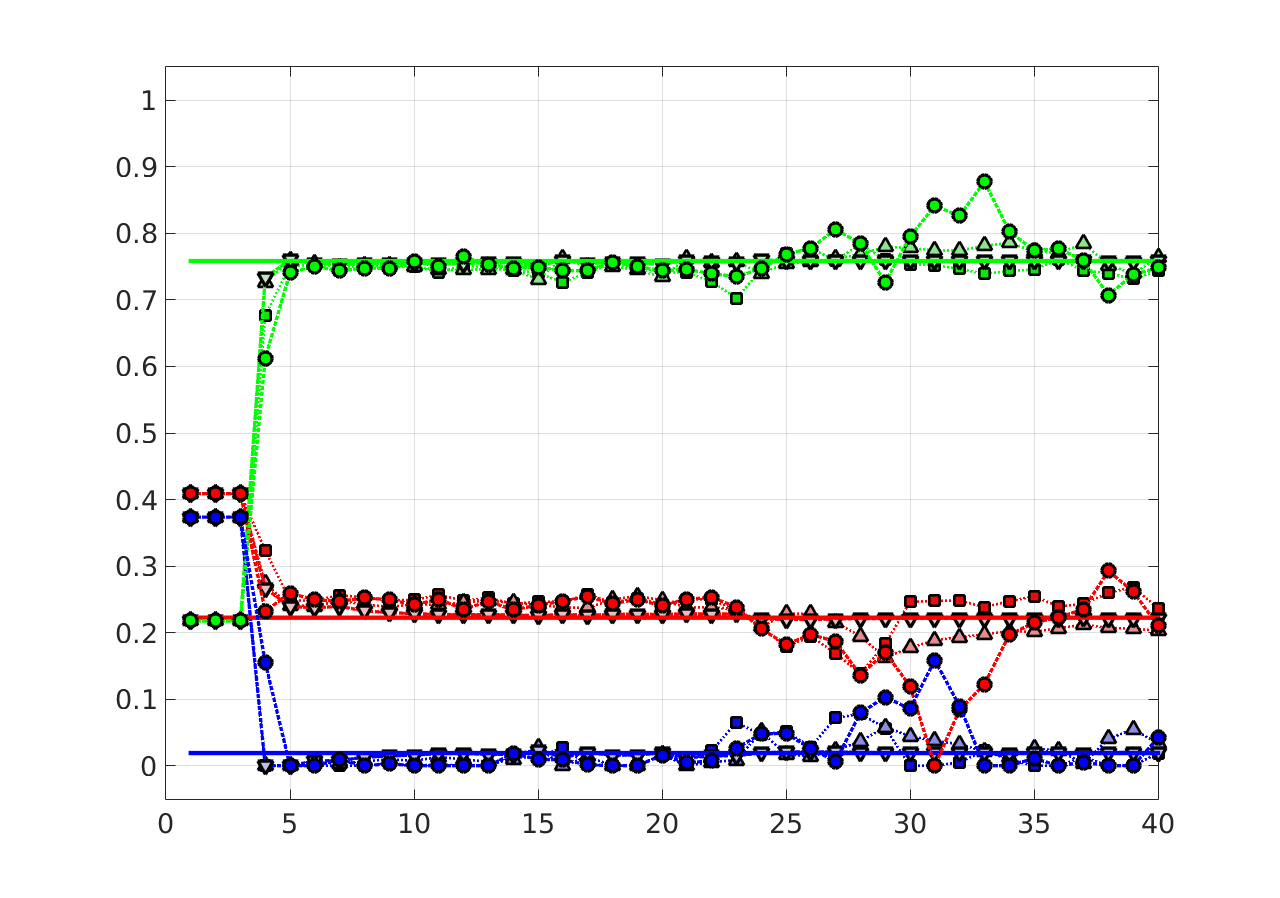}
 \caption{Estimated mixing parameters $\hat{\alpha}$ for $N=8$, $l=$, and noises variance  (\myinvertedtriangle{white} $S_w=0.0$, \mysquare{white} $S_w=0.1$, \mytriangle{white} $S_w=0.5$ and \mycircle{white} $S_w=1.0$)}
 \label{fig:estimation of alpha}
\end{figure}
Figure \ref{fig:decreasing cost with l 3D} shows the behaviour of the estimation error as a function of $l$ and $N$. This figure shows that the main factor in the reduction of the estimation error is the number of iterations $l$ used to update the estimates. It can be also seen that there a significant improvement in the initial iterations ($l < 10$), then after iterations there is no significant improvement in the estimation error. It is worth nothing that the error is decreasing with the iterations as it was shown in Section 4.

Figure \ref{fig:estimation of alpha} shows the time evolution of the estimated vector of mixing parameters $\hat{\alpha}$ for different values of process noise variance. The true values are representing as continuous line. When $S_w$ is smaller than the value of states ($S_w \leq 0.5$), the mixing parameters $\hat{\alpha}$ converge quickly to the true value or remain closer to it. 

\subsection{Example 2: Nonlinear time-varying system}
As a second example, we consider a second order time--varying nonlinear system whose dynamic is given by
behavior
\begin{equation}
  \begin{array}{rl}
     x_{k+1} &= \left[ \begin{array}{c}
         x_2\left(k\right)  \\
         p_1\left(k\right)x_1\left(k\right) + \sin(p_2\,x_2\left(k\right))
    \end{array} \right] + w_k   \\
   y_k &= C x_k + v_k
  \end{array}
\end{equation}
with the parameters $p_1(k)$ and $p_2(k)$ given by
\begin{equation}
  \begin{array}{rlr}
     p_1\left(k+1\right) &= 0.01 \, p_1 \left(k\right)\sin\left(\frac{5\pi k}{N}\right) & \hspace{0.5cm} \forall  k\in \mathbb{Z}_{\left[1,3N/4\right]},          \\
     p_1\left(k+1\right) &= p_1\left(k\right)        & \hspace{0.5cm} \forall k \in \mathbb{Z}_{> 3N/4}, \\
     p_2 &= 0.05
  \end{array}
\end{equation}
The polytope was designed to guarantee that the nonlinear system always remains inside it. The polytope is defined using three \emph{LTI} models
\begin{equation}        
  \begin{array}{c}
    A_1 = \left[ \begin{array}{cc}
         0 	& 	1.30  \\
         1 	& 	-1.52
    \end{array} \right],
    A_2 = \left[ \begin{array}{cc}
         0 	& 	-2.44  \\
         1 	& 	0.66
    \end{array} \right],
    A_3 = \left[ \begin{array}{cc}
         0 	& 	1.31  \\
         1 	& 	2.81
    \end{array} \right], 
    C = \left[ \begin{array}{cc}
    0.5 & 0.5 \\
    \end{array} \right].
  \end{array}
\end{equation}
%
%
The stage cost for all receding horizon estimators is chosen as $\ell(w,v) = w^T Q^{-1} w + v^T R^{-1} v$ with $R^{-1} = 5e^2$ and $Q^{-1}=diag\left(1e^3, 5e^3\right)$.
The proposed moving horizon estimator ($MHE_{A}$) the prior weighting matrix is given by $\Gamma_{k-N} \left(\chi\right) = (\chi - \hat{x}(0|k))^T P^{-1}_{k|k} (\chi - \hat{x}(0|k))$, where $P^{-1}_0 = 0.1 \boldsymbol{I}_2$ and $P_{k|k}$ is updated using equations \eqref{eq:updatePk} and \eqref{eq:updateXA} with $\sigma = 1$ and $c=1$. 
The robust moving horizon estimator ($MHE_{R}$) implements the algorithm proposed by \cite{muller2017nonlinear} with the nominal model, the prior weight given by $\Gamma \left(\chi\right) = L(\chi - \hat{x}(0|k))^T (\chi - \hat{x}(0|k))$ and parameters $\delta = 1$, $\delta_1 = \kappa^N$ ($\kappa=0.89$) and $\delta_2 = 1/N$ (see equation (3) of \cite{muller2017nonlinear}). 
The full information (\emph{FIE}, see \cite{ji2016robust}) is configured with the linearized model updated at each sampling time and the same parameters used by the $MHE_{ROB}$ with $\delta = 1$, $\delta_1 = \kappa^k$ and $\delta_2 = 1/k$. 
The robust Kalman filter was designed following the design procedure proposed by \cite{zhu2002design} using the nominal model and computing the bounds from the models of the polytope. The guess for the initial condition is $\bar{x}_0 = \left[0,0 \right]^T$, whereas $x_0 = \left[ 0.5, 0.3\right]^T$.

\begin{table}[!h]
  \centering
  \caption{Averaged MSE for $Sw = 0.1$, $Sv = 0.05$ and $N=8$.}
  \begin{tabular}{|l|c|c|c|c|}						\hline
    & \small$EKF_{R}$ & \small $FIE$ &\small $MHE_{R}$ &\small $MHE_{A}$        \\ 	\hline
    $x_0$ 	&\small0.40879 &\small0.32324 &\small0.31541 &\small0.018498			\\ 	\hline
    $x_1$ 	&\small0.4297  &\small0.30082 &\small1.0106 &\small0.064984 			\\ 	\hline
  \end{tabular}     \label{tab:average_estimation_error_for_nls}
\end{table}
Table \ref{tab:average_estimation_error_for_nls} shows the mean square estimation error ($MSE$) of each estimator averaged over 100 trials for $S_w = 0.1$, $S_v = 0.05$ and  $N=8$ for all receding horizon estimators ($FIE$, $MHE_{R}$ and $MHE_{A}$). It can be seen that the average mean square estimation error of $MHE_{A}$ is smaller than the other estimators ($EKF_{R}$, $MHE_{R}$ and $FIE$). 
The performance difference between the estimators that employ the nominal model ($EKF_{R}$ and $MHE_{R}$) is mainly due to the adaptation capabilities of $MHE_{A}$. The main performance difference between $MHE_A$ and $FIE$ estimators is due to the FIE attempts to reconstruct the state trajectory of a nonlinear time-varying system with a \emph{LTI} system.

\begin{figure}[b]
  \centering
  \includegraphics[width=0.90\textwidth]{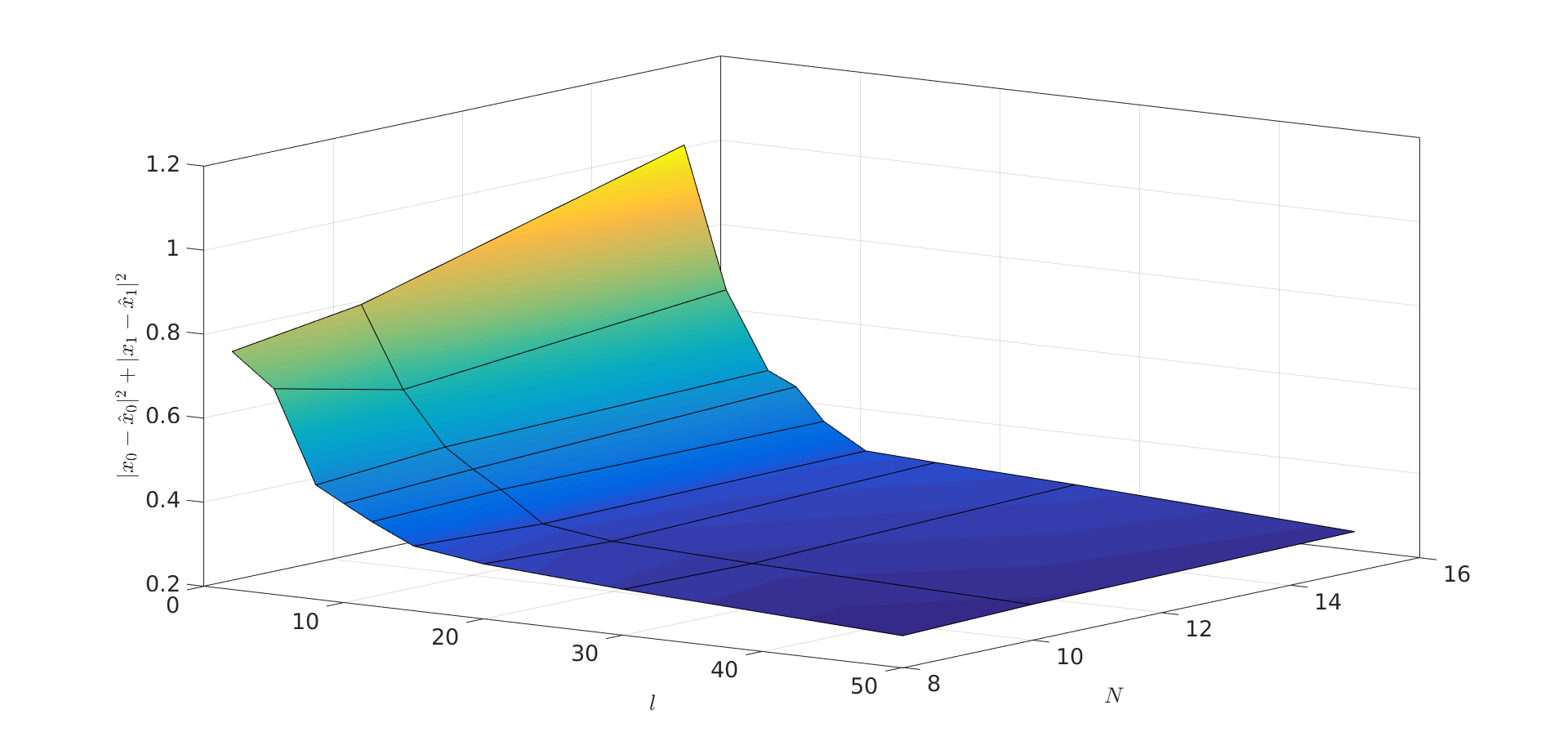}
  \caption{Performance of the proposed algorithm for different values of $l$ and $N$.}
  \label{fig:error NLS Nvsl}
\end{figure}
Figure \ref{fig:error NLS Nvsl}  shows the behaviour of the estimation error as a function of $l$ and $N$. This figure shows that the main factor in the reduction of the estimation error is the number of iterations $l$ used to update the estimates. It can be also seen that there a significant improvement in the initial iterations ($l < 15$), then after iterations there is no significant improvement in the estimation error. 
It can also see how the estimation error increases for higher values of $N$. This behavior is due to the estimator use only one model along the entire estimation horizon, whereas the nonlinear system is changing its parameters every sample.
The $EKF$ aim to improve the estimation error in comparison with the $MHE_{A}$. This improvement is due to the $EKF$ update the model at every sampling time with the true model. Besides, the $MHE_{A}$ use the same model along the entire horizon estimation to estimate the optimal state trajectory of the nonlinear system.

\begin{figure}[tb]
	\centering
	\includegraphics[width=0.90\textwidth]{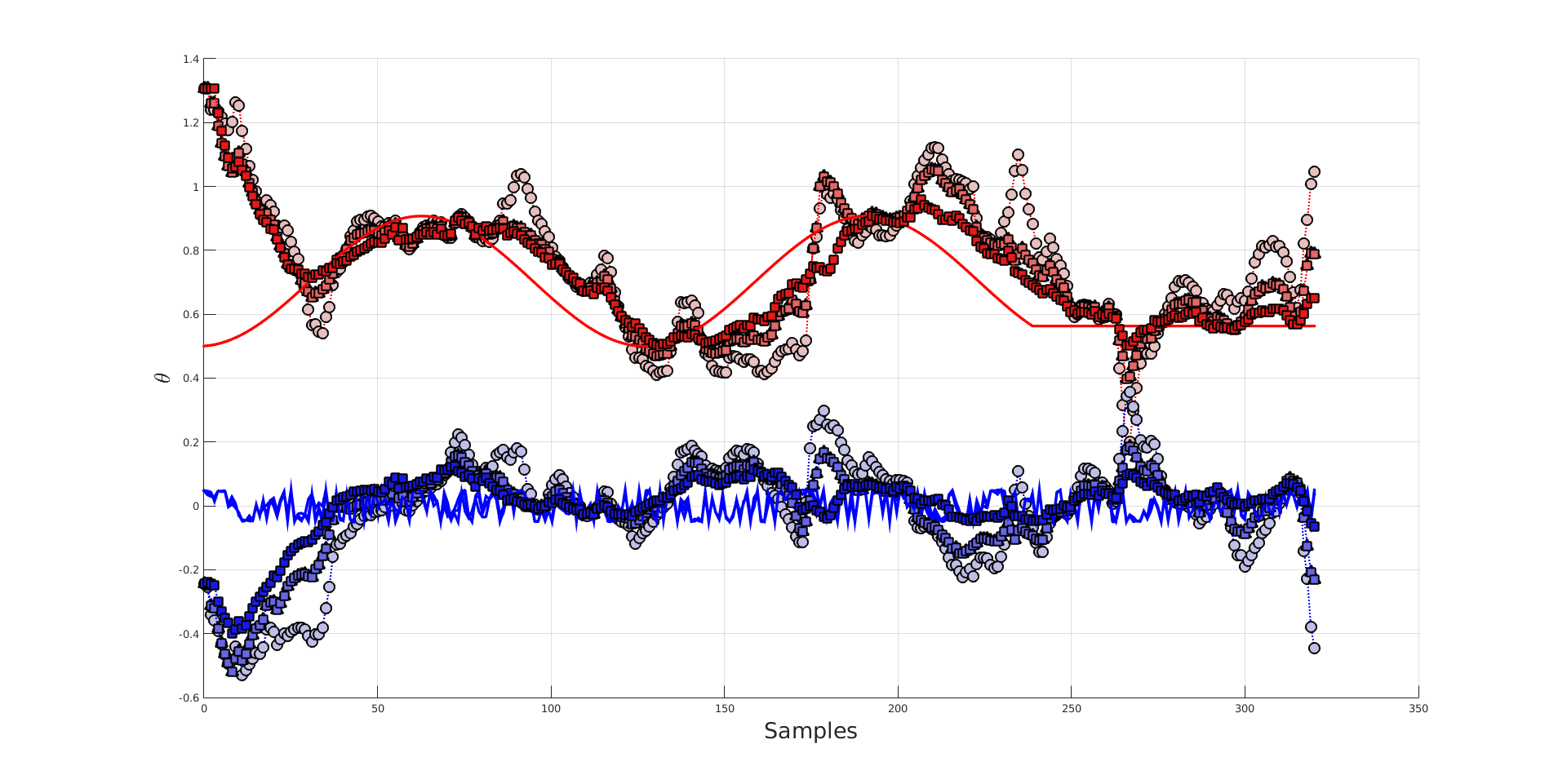}
    \caption{Estimation of the parameters of the nonlinear time-varying system ( $N=8$, and noises \mysquare{white}=0.1, \mytriangle{white}=0.25 and \mycircle{white}=0.5).}
    \label{fig:parameters_estimated_NLS}
\end{figure}
Figure \ref{fig:parameters_estimated_NLS} shows the time evolution of the estimated parameters $\hat{p}_1$ and $\hat{p}_2$ for different values of process noise variance. The true values are representing as continuous line. When $S_w$ is smaller than the value of states ($S_w \leq 0.25$), the parameters $\hat{p}$ converge quickly to the true value or remain closer to it. However, 


\section{Conclusions}

In this paper we introduce an adaptive polytopic observer for nonlinear systems under bounded disturbances based on moving horizon estimator and dual estimation techniques and proved their stability properties. In a first stage we proved the stability of the dual estimation iteration. Then, in a second stage we established robust
asymptotic stability for the adaptive moving horizon estimator. It was also shown that the estimation error converges to zero in case that disturbances converge to zero.

An advantage of this updating mechanism is that the required conditions on prior weighting are such that it can be chosen off-line. Furthermore, it introduces a feedback mechanism between the arrival cost weight and the estimation errors that automatically controls the amount of information used to compute it, which allows to shorten the estimation horizon.


\begin{ack}                               
The authors wish to thank the Consejo Nacional
de Investigaciones Cientificas y Tecnicas (CONICET)
from Argentina, for their support. 
\end{ack}

\bibliographystyle{agsm}
\clearpage
\bibliography{autosam.bib}       


\end{document}